# Model Choice and Crucial Tests. On the Empirical Epistemology of the Higgs Discovery

Peter Mättig[1] & Michael Stöltzner[2]
-

*Abstract*: Our paper discusses the epistemic attitudes of particle physicists on the discovery of the Higgs boson at the Large Hadron Collider (LHC). It is based on questionnaires and interviews made shortly before and shortly after the discovery in 2012. We show, to begin with, that the discovery of a Standard Model (SM) Higgs boson was less expected than is sometimes assumed. Once the new particle was shown to have properties consistent with SM expectations – albeit with significant experimental uncertainties –, there was a broad agreement that 'a' Higgs boson had been found. Physicists adopted a two-pronged strategy. On the one hand, they treated the particle as a SM Higgs boson and tried to establish its properties with higher precision; on the other hand, they searched for any hints of physics beyond the SM. This motivates our first philosophical thesis: the Higgs discovery, being of fundamental importance and establishing a new kind of particle, represented a crucial experiment if one interprets this notion in an appropriate sense. By embedding the LHC into the tradition of previous precision experiments and the experimental strategies thus established, Duhemian underdetermination is kept at bay. Second, our case study suggests that criteria of theory (or model) preference should be understood as epistemic and pragmatic values that have to be weighed in factual research practice. The Higgs discovery led to a shift from pragmatic to epistemic values as regards the mechanisms of electroweak symmetry breaking. Complex criteria, such as naturalness, combine epistemic and pragmatic values, but are coherently applied by the community.

## 1. Introduction[3]

The discovery of a[4] Higgs boson at the Large Hadron Collider (LHC) of the European Laboratory CERN, announced in July 2012, is arguably one of the most important scientific achievements of the past few decades. The discovery received world-wide attention; two of the inventors of the Higgs mechanism, François Englert and Peter Higgs, were awarded the 2013 Nobel Prize in physics. The Higgs boson had been the final piece of the so-called Standard Model of particle physics (SM) not observed by previous experiments. The Higgs mechanism in the SM was required to generate masses of the elementary particles in a consistent way. Even though with the discovery of the Higgs boson, the SM – terminology notwithstanding – has now become one of the most successful scientific theories of contemporary physics, all particle physicists agree that it will not be the final word. There are both compelling internal and external reasons to postulate physics beyond the SM (BSM).

---

[1] Bergische University of Wuppertal, Department of Mathematics and Natural Science, Gaußstrasse 20, 42119 Wuppertal; now at Department of Physics and Astronomy, University of Bonn, Nussallee 12, 53115 Bonn, Germany; Email: peter.mattig@cern.ch.
[2] Department of Philosophy, University of South Carolina, Columbia, SC 29208, USA; Email: stoeltzn@sc.edu.

[3] The study was performed as part of the project 'Model Dynamics' supported by the DFG (project no. MA 2793/2-1). It is based on questionnaires that were developed by the authors, Arianna Borrelli, Robert Harlander, and Friedrich Steinle, and interviews conducted by Arianna Borrelli. Karsten Egger assisted in the evaluation of the questionnaire. We acknowledge the help of Annette Holtkamp in obtaining the SPIRES email list. We acknowledge discussions within the recently established DFG Research Unit 'Epistemology of the LHC' and detailed comments by Robert Harlander, Martin King, and Gregor Schiemann. Some other results from this study can be found in (Borrelli 2016) and her presentation at CERN available under https://indico.cern.ch/event/232108/. We thank the anonymous referees for their most helpful criticism and manifold suggestions. We thank Sophie Ritson for having made us aware of Baetu (2017).

In retrospect, the discovery of the Higgs boson might seem to be just the final step in a long series of discoveries and precision tests in which stronger and stronger accelerator experiments confirmed all particles of the SM and scrutinized their interactions. The present paper argues that as regards the community of elementary particle physics this picture needs qualification. In actual fact, even shortly before the Higgs discovery a significant percentage of physicists raised concerns whether it would at all be found at the LHC and expressed preferences for other explanations of the particle masses.

In this paper, results of questionnaires and interviews with LHC physicists shortly before (autumn 2011) and shortly after (autumn 2012) the discovery are presented and analysed. From these empirical sources, we reconstruct the physicists' beliefs in the adequacy of certain models, in the outcome of the LHC experiments, and concerning the possible impacts of the LHC experiments on those models. This will help us to understand the epistemic attitudes of particle physicists, and the principles and strategies guiding their research. Our empirically informed epistemological investigation also promises new insights for a philosophical analysis of how actual and expected experimental findings, on the one side, and pragmatic quality criteria of models, on the other, influence the research agendas of particle physicists.

We have limited the scope of the present paper to a specific part of the empirical material available in the questionnaires and interviews: to the Higgs mechanism and competing accounts of mass generation, and to the presently most discussed pragmatic quality criterion, naturalness. It must be said, however, that the LHC was, from the very beginning, designed not only to search for the Higgs boson but also to probe the deep TeV energy range and find signs of BSM physics. Whereas the first objective has now been achieved, no 'new physics' BSM has been observed to date.[5]

The specific descriptive questions addressed in this paper are as such:

1. Did physicists in 2011 expect the Higgs boson to be discovered at the LHC and how did they evaluate the Higgs candidate in 2012, that is, before its properties were known to a sufficient extent? What was their assessment of alternative models for mass generation in 2011 and in 2012?
2. How important was the naturalness problem, a major guiding principle to develop models of physics beyond the SM (BSM), in shaping physicists' attitudes and preferences?

We will show that, in 2011, physicists were rather undecided whether the SM Higgs boson would eventually be found, that is, even a few months before the first evidence was reported. However, once a candidate had been observed in 2012, they quickly embraced the notion that 'a' Higgs boson had been found. Its discovery immediately affected the research directions in particle physics. The experimental results pulled in different directions as regards the naturalness problem. There was, on the one hand, less motivation to search for alternatives to the Higgs mechanism. On the other hand, after finding the Higgs boson, the naturalness problem posed by the scalar Higgs particle changed from a virtual into a real problem, that is, there existed empirical results directly relevant for it. But since 2012 no BSM effect to cure this problem has been found. This has led some physicists to develop a more critical attitude as to naturalness' significance for elementary particle physics.

---

[4] We adopt the usual terminology and address the SM Higgs as 'the' Higgs, whereas those models with a potentially more complicated Higgs sector as containing 'a' Higgs boson.
[5] The attempts at finding physics BSM and their effects on the thinking of LHC physicists will be discussed in a separate paper.

The physical developments prompt the following philosophical questions.

3. What do the epistemic attitudes of particle physicists shown in the questionnaire and the interviews mean for the significance and application of criteria of theory (or model) [6] choice and the principles and epistemic values guiding model development?
4. What does the comparison of the situations before and after the discovery of the Higgs boson signify for the relationship between theory (or models) and experiment? In particular, was the Higgs discovery a crucial experiment for the SM?

The paper is organised as follows. After a brief introduction into the theoretical motivation for the Higgs mechanism and the experimental attempts to find evidence for a Higgs boson (Section 2), we provide the background of the philosophical problems raised (Section 3) and discuss the methodology of our study (Section 4). The presentation of the results will be subdivided into the outcomes of the questionnaire and the interviews in 2011 (Section 5) and in 2012 (Section 6) respectively. Finally (Section 7), we outline our answers to the above-mentioned four questions.

## *2. The physics of electroweak symmetry breaking*

Several articles of both physicists and philosophers discuss the emergence of what is by now called the 'Higgs' mechanism (Cf. Ellis, Gaillard, Nanopoulos, 2015; Nobel laudatio 2013; Karaca 2013b). Here, only a brief account of the motivation and the concepts behind the Higgs boson can be given. In the early 1960s, various models were developed to unify two interactions governing the subnuclear world, the electromagnetic and the weak ones. These unifications adopted the concept of local gauge symmetry that had previously been applied successfully to quantum electrodynamics (QED). In brief, this symmetry means that the theory is invariant under a specific space-time dependent transformation of the quantum fields. Assuming this symmetry in the SM leads to a consistent interacting field theory, which for particle physicists means free of any infinities after renormalization.

However, a major problem that physicists were facing in applying local gauge symmetry to weak interactions was that observations implied that the corresponding gauge bosons have a non-vanishing mass. As such, gauge boson masses break the symmetry explicitly, thus leading to theoretical inconsistencies, such as the violation of unitarity. To remedy this, in the 1960s, physicists used the concept of spontaneous symmetry breaking (SSB) to generate gauge boson masses in a gauge invariant way at the cost of introducing an additional scalar, i.e. spin-less, particle, which became known as the Higgs boson. This particle was discovered at the LHC some 50 years after its invention. The Higgs sector of the SM is a novel element in physics, in that it describes the mass of elementary particles in terms of their interaction with an elementary scalar field.

---

[6] In line with the current philosophical literature, we consider models as autonomous entities in scientific theorizing, not as the logical models of a theory. For how one can apply this conception to elementary particle physics, cf. Borrelli & Stöltzner (2013) and Stöltzner (2014).

Whereas the weak and the electromagnetic components of the 'electroweak' theory have almost the same strength at very high energies, they are substantially different at low energies, since only the weak interaction invokes a massive interacting particle. Therefore, the mechanism of mass generation is also referred to as 'electroweak symmetry breaking' (EWSB). The Higgs mechanism was originally only devised to give mass to the weak gauge bosons $W^{+/-}$ and $Z^0$. (The latter represents the electrically neutral component of the weak interaction, which, however, has an admixture of an electromagnetic component.) It turned out that the Higgs mechanism could also be applied to give masses to fermions, through a Yukawa interaction, albeit without predicting their numerical values.

### 2.1 The experimental search for the Higgs boson

The general conception of the Higgs mechanism just outlined was developed into phenomenological predictions[7] opening the way for experimental searches of the Higgs boson. Given the masses of the *W* and *Z* bosons, the Higgs mechanism introduced just one additional parameter to the SM that had to be determined by experiment, notably by measuring the Higgs mass[8]. Whereas the theory did not provide a prediction for this mass, it did lead to an upper bound of 800 GeV to maintain theoretical consistency. Depending on its mass, it could be unambiguously predicted how the Higgs boson is produced and the way it can be seen by experiments. Since a Higgs boson would only exist for small fractions of a second, it would decay, depending on its mass, mainly into massive fermions and *W* and *Z* bosons.

As a result, a clear strategy for finding the Higgs boson was devised.[9] However, this did not make Higgs searches easy. Essentially no experiment before the start of CERN's Large Electron Positron Collider (LEP) in 1989 was sensitive to the Higgs boson. At the end of the LEP data taking no significant signal was observed. However, the sensitivity of LEP was such that a Higgs of 114.4 GeV or less should have been found, allowing physicists to place a lower limit on the SM Higgs mass. Between the end of LEP and the start of LHC, an additional small mass interval around 160 GeV could be excluded at the Tevatron. The outstanding precision of the LEP data and theoretical calculations based on the SM provided an indirect sensitivity to the Higgs mass by quantum fluctuations, e.g. loop corrections to the W and Z bosons, bounding it to be lighter than 157 GeV.

In 2010, data taking at the LHC started for the final assault. It was clear that the LHC had the sensitivity to observe the Higgs boson in the remaining allowed mass range, using the decay modes that were unambiguously predicted for a SM Higgs. Relatively soon one could exclude a high mass Higgs of 200-600 GeV – in full agreement with LEP's indirect limits. With the rapid increase in data rate, both the ATLAS and CMS experiments reported, at a CERN colloquium on December 13, 2011, an excess of events that could be taken as initial evidence for a new particle around 126 GeV. On the other hand, the probability that this would be just a background fluctuation was still too high to claim an observation. However, half a year later much more data had been accumulated, such that both detectors presented, at a special CERN seminar, a signal of 5 standard deviations each. The data correspond to a background fluctuation probability of about $10^{-9}$, where the background is considered as SM without Higgs. By convention in particle physics, this was sufficiently small to claim a discovery, an observation.[10] A few weeks later, the two experiments published their data. (Aad et al. 2012, Chatrchan et al. 2012).

---

[7] Cf. Ellis, Gaillard, Nanopoulos (1976).

[8] More precisely, the Higgs potential has two parameters, one of which is related to the masses of the W and Z – bosons and connected to the vacuum expectation value v leaving one additional parameter that had not been determined before the Higgs discovery. It should be noted that in the SM the Yukawa couplings g, i.e. the Higgs couplings to the fermions are given by g=sqrt(2)*$m_f$/v, $m_f$ being the mass of the fermion. Therefore, the Yukawa couplings re-express the fermion masses but are not counted as extra free parameters.

[9] One publication was even titled "The Higgs Hunter's Guide" (Gunion et al. 1990).

Still, the data were not sufficient to definitely claim this to be the long-awaited Higgs boson. Some important properties had not yet been confirmed, and the precision of the measurements on production and decay properties was still marginal. On the other hand, those properties that were observed corresponded to what is expected for a SM Higgs boson. For instance, the particle had been found in two decay modes with rates consistent with the expectation, and it had a mass in agreement with the direct and indirect limits known from previous experiments. As of today (2017), more properties of the discovered particle have been studied, the decay modes and the mass have been measured to higher precision, in accordance with the SM. Even though there is still need for further measurements, the majority among physicists now considers the new particle is indeed the Higgs boson.

### *2.2 Alternatives to the Higgs boson*

Already shortly after the invention of the Higgs mechanism, several authors expressed discontent because this solution of the SSB problem appeared largely ad-hoc. For example, it has limited predictive power in that it cannot determine the quark and lepton masses. Concern was also raised that the Higgs mechanism introduces a new concept into the theory for the one and only purpose of mass generation. Over the years, the list of issues cited by physicists in this respect has expanded. (Cf. Friederich, Harlander, & Karaca 2014, sect. 3).

Several alternative mechanisms of EWSB have emerged over the past decades. They used a scalar particle and a Higgs-like potential to generate mass. However, in many cases, the conceptual framework of the alternative models was very different from, and implied physics beyond the SM. These BSM models will be considered in this article only in relation to mass generation.

A fairly straightforward modification of the original Higgs mechanism was to extend the Higgs sector. Originally, one complex Higgs doublet was assumed, leading to four fields one of which would be the observable Higgs boson, whereas the others would not be directly observable. However, one can also introduce, e.g., a second doublet leading to five physical elementary Higgs bosons with no change in the principal mechanism of mass generation. Such models allow the different Higgs bosons to assume different roles.

The two Higgs doublet model is of special interest in BSM considerations since it is the minimally required Higgs sector in the framework of Supersymmetry, the most often discussed extension of the SM. Supersymmetry assumes a new fundamental symmetry of particles with integer and half-integer spins. In the context of LHC physics,

---

[10] On the criteria when particle physicists claim ‚evidence', versus ‚observation' or ‚exclusion' see the 'Prologue' to Franklin (2013).

Supersymmetry typically is studied in several variants of the MSSM ('minimal supersymmetric SM') that contain a rather broad range of features that allow one solve some basic problems of the SM. Among those is providing a Dark Matter candidate and solving the naturalness problem (see below). Moreover, the MSSM is the only BSM model that makes a firm prediction on the upper limit of the Higgs mass: it has to be lighter than about 130 GeV, a limit that is much tighter than the range allowed by the SM. (This limit of 130 GeV is valid for all SUSY models considered at the LHC and for SUSY scales of some 1 TeV.)[11]

Another class of models assumes the Higgs boson to be a composite, i.e. made up of sub-constituents. The first model of this kind was devised at the end of the 1970s by essentially copying concepts known from the strong interactions that explain hadron masses. This mechanism was dubbed 'Technicolour'; since it involved strong interactions, it was considered as a type of 'strong' or 'dynamical' EWSB. The realisation of these models led to inconsistencies with measurements, such that this approach by now has become disfavoured. However, the concept of composite Higgs particles has been implemented within multiple frameworks invoking additional symmetries, new interactions, or additional spatial dimensions (e.g. Csaki and Tanedo 2016).

All these alternative models assume scalar particles like the SM Higgs boson to generate the masses of gauge bosons and fermions. However, the properties of these scalars are different, albeit sometimes by a rather small amount given by tuneable free parameters. All of them also lead to new phenomena, e.g. more scalars and more fermions.

## *2.3 The Naturalness problem*

From a theoretical perspective, the existence of an elementary scalar Higgs boson introduces an 'unnaturalness' into the SM. The concept 'naturalness' was introduced in slightly different forms by 't Hooft (1979) and Susskind (1979). The problem itself has a longer history (cf. Giudice 2008) and reaches beyond the context of the Higgs problem (cf. Giudice 2013). During the past decades, naturalness has developed from a merely technical problem into an influential guiding principle for BSM physics; that is, extensions of the SM were developed with the explicit aim to remedy the naturalness problem.

In a nutshell, the naturalness problem is this: since the fundamental equations of the SM can only be solved in a perturbative expansion, at each order a theoretically well-defined correction has to be applied to compensate for quantum fluctuations that would modify a physical quantity like mass or charge. Such 'renormalisation' is a standard technical procedure in theoretical particle physics. For the SM particles of spin ½ or 1 these corrections are of a few percent. In the case of the Higgs boson, which is a scalar, however, the correction to the square of the Higgs mass grows quadratically with energy.

Introducing a cut-off mass where the theory would break down, leads to finite corrections. In the case of the SM, this could be at the rather high Planck scale, where gravity becomes important and the SM is known to be insufficient. Assuming such a scale within the SM, in case of the Higgs mass, makes these corrections appear 'dramatic and even bizarre' (Peskin and Schroeder 1995, p. 788); for instance, in order to keep the square of the Higgs mass at its measured value of 125 GeV, corrections have to be invoked that are more than $10^{30}$ times higher than the Higgs mass itself. Furthermore, these corrections have to be fine-tuned over many decimal places. Although theoretically viable and consistent, the magnitude of these corrections is considered 'unnatural'. Once this correction is defined the theory is completely consistent and any dependence on the scale is eliminated.

---

[11] There exists a small (logarithmic) dependence of this bound on the masses of the SUSY particles. Even if more Higgs multiplets exist, the bound would only rise to 150 GeV.

During the past two decades, naturalness has arguably become the most influential guiding principle for constructing and motivating BSM models. Or more specifically, many physicists believed that if a SM Higgs boson existed, it would come with new phenomena to keep the theory 'natural'. For instance, new symmetries, extra spatial dimensions, or a composite Higgs boson built from smaller objects would avoid unnaturalness. Allowing for corrections of just a few percent – as for the other sectors of the SM – these new phenomena should be in the mass range of 1 TeV that is well covered by the LHC. One has to be aware that there is no clear definition of when a theory would become unnatural and there is a large freedom how much fine tuning is considered acceptable. Yet once a bound on the acceptable fine tuning is set, it determines the mass range at which new phenomena are expected. At any rate, thus far there has neither been a direct observation nor any clear indirect indication from precision studies that such a new effect exists.

## 3. Philosophical Background: Theory Choice and Crucial Experiments

Our empirical study allows us to address two longstanding problems in philosophy of science from the perspective of the actual practice of scientists. First (in 3.2), we discuss the relationship between epistemic and pragmatic (including aesthetic) criteria of theory choice in the contexts of models of electroweak symmetry breaking. Presently most discussed among these criteria is naturalness. Second (in 3.3.), we discuss under which conditions complex experiments, such as the Higgs discovery, are considered decisive or even crucial. We begin this section, however, by showing that the present debates about naturalness represent a case in point about the influence of criteria of theory choice. The general aim of the present section is to give a short survey of the current philosophical discussion that provides the basis for Section 7.

### 3.1. The Philosophical Challenge of Naturalness

Several facets of the naturalness problem have attracted philosophers' attention; among them are its precise content and to what extent it influences current research in particle physics. Porter Williams (2015) has distinguished four (closely related) ways to formulate the naturalness problem: (i) quadratic divergences in renormalisation; (ii) 't Hooft's (1979) suggestion that setting a small parameter to zero must increase the symmetry of the system; (iii) a specific version of the problem of fine-tuning of fundamental constants; (iv) an aesthetic criterion, whose force is derived from various factors prevailing within the scientific community. Williams argues that none of his four reformulations captures the whole naturalness problem and believes that it is rather an expression of the central dogma of effective field theories according to which widely separated scales should eventually decouple.

The physicist James Wells (2015) considers (i) as the root of the problem, but subsequently emphasizes the significant difference between the technical naturalness (ii) and the absolute naturalness involved in fine-tuning that eventually goes back to Dirac's classical worries about large dimensionless numbers. He elaborates an example of an exotically augmented quantum electrodynamics (QED) that consistently instantiates absolute naturalness at the expense of "more parameters, more fields, and more complexity in the theory." (2015, 107) He admits that this principle is controversial, but believes, more generally, "that in the era of the Standard Model's ascendancy, the influence of simplicity and Ockham's razor to theory construction has paled in comparison to Naturalness." (2015, 104) [12]

Grinbaum (2012) instead has argued that – in virtue of its complex nature – naturalness is exclusively an aesthetic criterion. Williams (2015) rejects Grinbaum's interpretation because aesthetic criteria are notoriously ambiguous. Supersymmetry, for instance, is considered most promising by many physicists, even though it is aesthetically attractive in the unbroken state but aesthetically unattractive after its breaking produces a large number of new constants. Borrelli (2015) argues, that it is precisely the vagueness of the concept of naturalness that allows it to function as a useful common narrative of the different subcultures of particle physics, the experimentalists and theoreticians

The goal of the present paper is not to analyse all facets of naturalness. Instead we take it as the currently most important example of a guiding principle for a 'good' model within contemporary particle physics and provide empirical results about its relationship with other guiding principles. More specifically, we will compare the relatively new and quantitative concept of naturalness with the more familiar pragmatic, aesthetic, and qualitative criteria of elegance and simplicity – Ockham's razor being one of its manifestations.

*3.2 Epistemic and pragmatic criteria of theory choice*

Philosophers have traditionally distinguished epistemic and pragmatic criteria of theory choice (or preference). The former, among them empirical adequacy and theoretical consistency, are held to be rationally compelling. Pragmatic criteria have instead been seen as a way to decide among epistemically equivalent alternatives by appealing to a theory's simplicity or other aesthetic features, or to its fruitfulness for further research. Among the classical examples are the choice between a geocentric and a heliocentric world view at the time of Copernicus and the early philosophical debates about the nature and alleged conventionality of space and time. The philosophical significance of these criteria of theory choice arises from the problem of underdetermination of theory by empirical evidence that Pierre Duhem illustrated at the parallelism between Newton's corpuscular theory and Huygens's wave theory of light. Duhem argued that experimental data never uniquely determine a particular hypothesis because setting up and confirming a hypothesis presupposes the correctness of many other hypotheses including the theories governing the measurement devices. If one accepts some version of the underdetermination argument, pragmatic and aesthetic criteria become more relevant or even inevitable.

Underdetermination is also discussed under the rubrics of theory-ladenness of data or – following Neurath and Quine – confirmational holism. This means that any experimental result confirms or refutes both the theory or model under investigation and a large set of other assumptions that are assumed to be true. Especially in Quine's hands, underdetermination and holism took a logical and semantic tack that not only ruled out that empirical evidence could deductively entail scientific theories, but that additionally seemed to imply that any theory could be rationally retained in the face of recalcitrant evidence. Laudan has pointed out that, while the latter may be logically possible, scientists do not act "in an evaluative vacuum." (1990, 276) To his mind, the non-uniqueness of theory resulting from Duhemian underdetermination can be accepted without adopting an egalitarian approach towards rival theories. Laudan and Leplin, more generally, held "that the epistemic bearing of evidence on theory is ... subject to reinterpretation as science grows and may be indeterminate at a particular point in the process of growth." (1991, 455) Norton, moreover, has argued that ampliative inferences remain valid if underdetermination focuses locally, "on the confirmation of hypotheses by scientists in actual scientific practice," (2008, 23) rather than being taken as a global challenge to its rationality. But overall the topic remains controversial (cf. Stanford 2017). Since the present paper is concerned with the analysis of experimental and theoretical practice, we are following Laudan in focusing on the scientific-practical aspects of underdetermination This focus on scientific practice is also a better basis for assessing the role of pragmatic and epistemic criteria of theory choice than debates about the rationality of science globally.

---

[12] Ockham's razor instructs us not to add basic entities without any need to do so. It has originally been a metaphysical principle, but the term is nowadays used more broadly. Cf. the following section; see also Wells (2017).

Speaking of theory choice, philosophers of science have traditionally set pragmatic criteria firmly apart from the epistemic criterion of empirical adequacy and all other scientific questions that can be resolved within an explicitly formulated theoretical framework (cf. Carnap 1950). Thomas S. Kuhn (1977) rejected this separation and advocated a broader list of characteristics of a good scientific theory. It includes: empirical "accuracy, consistency [internally and with respect to other theories], scope, simplicity, and fruitfulness." (1977, 322) These five criteria of theory choice are not mutually independent; they are often context-dependent and may point in opposite directions. For instance, an increase in accuracy can trivially be obtained by adding additional parameters; yet scientists may prefer to make ado with a smaller number of fundamental quantities – or with a simpler law – even at the expense of some accuracy. Thus, scientists have to assess the relative weight of these criteria when deployed together. Both their form and the relative weight, to Kuhn's mind, contain contextual and idiosyncratic (psychological) factors. Kuhn was however at pains to argue that such subjectivity does not render theory choice irrational or a mere matter of taste. Theory choice, we might add, was not a major battle in the conflict between historical rationality and historical contingency waged during the 1970s. Kuhn's point was the historical and factual nature of theory choice, not its contingency or arbitrariness. Historians often find an increasing unanimity of individual choices in a certain field. Such factual unanimity does not establish rationally binding criteria for theory choice. Instead of being rules of an algorithm, the criteria of theory choice function "as values, which influence it. ...; they do specify a great deal: what each scientist must consider in reaching a decision." (1977, 331)[13]

Heather Douglas has proposed a finer-grained account in order to restore the separation between epistemic and pragmatic cognitive values and reduce conflicts between them. She distinguishes (i) minimal criteria applied to the theory per se, among them internal

---

[13] Laudan reads Kuhn's analysis of theory choice against the backdrop of scientific revolutions that represent breaks in rational justification. This rehearses, to Laudan's (1990) mind, the holistic and egalitarian reading of underdetermination and provides a justification for the sociologizing of epistemology. Without entering into a broader Kuhn debate, it seems to us that once we limit ourselves to an epistemic or local understanding of underdetermination, the Kuhnian analysis of the values of a good scientific theory can still provide important insights into scientific practice. As Kuhn himself has emphasized, these values are only one element of theory choice, alongside sociological factors and inductive reasoning. Moreover, our goal here is not to find all determinants of theory choice, but to focus on the role of the epistemic and pragmatic criteria or values in the preference of models in elementary particle physics.

consistency; (ii) minimal criteria applied to the relation of theory and evidence, among them empirical adequacy; (iii) desiderata applied to theories per se, among them scope, simplicity, and potential explanatory power of a theory that largely "fall under the rubric of the fruitfulness of the theory". (2013, p. 800); (iv) desiderata applied to the relation of theory and evidence, among them being supported by a broad range of empirical evidence and not being contrived to match a small domain of facts in an ad hoc fashion. While the values in categories (i) and (ii) are epistemic, category (iii) contains "strategic or pragmatic values" (2013, p. 800) that help in "deciding which theory to pursue next" (2013, 804)). Instead, group (iv) "provides assurance that our scientific claims are more likely to be reliable." (2013, p. 800) Moreover: "While simplicity, scope, and explanatory power are often thought to pull against each other when considering theories alone (group iii), they pull together when considering a theory in relation to evidence (group iv)." (2013, 803)

Perhaps, the most important pragmatic criterion in the history of particle physics is simplicity. Most influential has been the quest for a simple unified theory of all fundamental forces.[14] Simplicity also stands behind particle physicists' long-time worries about the many parameters that are needed to make the SM empirically adequate. As Baker (2013) rightly observes, it is quite challenging to pin down the notion precisely. Many authors distinguish elegance (typically attributed to a theory) and parsimony (Ockham's razor that directs us not to introduce unnecessary entities). Both aspects of simplicity may come into conflict. For instance, the introduction of supersymmetric partners to all fundamental particles reduces the basic components into chiral super multiplets, thus reducing the complexity of the theory. The elegance of an exact symmetry between fermions and bosons in the unbroken theory disappears once a breaking mechanism is introduced, which leads to a large number of additional parameters.

From the interviews and questionnaires, we will analyse in Sect 7.3 and 7.4 how particle physicists understand and weigh epistemic and pragmatic values and how they assess the criterion of naturalness in BSM models. Applying the philosophical debate about those values to model preferences within a variegated model landscape has certain consequences on how to interpret such preferences further. We are following Kuhn and Douglas in speaking about values rather than criteria, and will also speak about preference instead of choice even in cases, such as supersymmetry or not, where the latter terminology could be appropriate.

### 3.3 Making experiments crucial

The second classical philosophical problem relevant for the present paper concerns the interaction between theory (or models) and experiment. LHC's first task consisted in a definitive and crucial test of the SM, i.e. to find the Higgs boson or exclude its existence. Since the Higgs boson is an essential part of the SM and since LHC would cover the whole energy scale relevant for direct searches, not finding it should have eventually implied that the SM was refuted. Thus, a large majority of elementary particle physicists interviewed expressed the conviction that a Higgs discovery or non-discovery at LHC represented a crucial and decisive test for the SM.

---

[14] Note that some philosophers – and some physicists, perhaps – would argue that there are metaphysical reasons or some a priori principle of rationality that imply that a simpler theory is more likely to be true. Such questions are, however, outside the scope of the present paper.

This widely shared conviction among physicists prompts the question whether the Higgs discovery represented a crucial experiment in a philosophical perspective? Let us take a closer look. The term 'crucial experiment' originated with Francis Bacon and became influential through Newton and his demonstration that sunlight consisted of rays exhibiting different behaviours. A crucial experiment, in this traditional understanding, unambiguously and definitively confirms a hypothesis or decides between rivalling hypotheses. Pierre Duhem objected on the basis of the underdetermination argument.

This philosophical context has made scholars wary about crucial experiments, especially if they understood underdetermination as a primarily logical and global problem and followed Duhem in allowing only deductive inferences between theories and data.[15] While some emphasized that falsifications of a theory were more likely to be crucial experiments than corroborations, Lakatos famously objected to this asymmetry and bluntly stated: "No experiment is crucial at the time it is performed (except perhaps psychologically)." (1974, 320) His main argument was that the assessment of each experiment can only be performed against the backdrop of the entire research program it is embedded into and against its competitors. Thus, designating an experiment as crucial is partly a historical assessment.

The idea that a crucial experiment is embedded into a broader program is also the core of a recent debate about crucial experiments in biology. Weber (2009) defends the characterization of an experiment as crucial, not within the traditional contexts of deductive reasoning and the refutation of alternative hypotheses, but by developing "an experimentalist version of inference to the best explanation." (2009, 21) Hypotheses are not refuted, but positively selected as those best supported by the evidence. Weber's strategy to defend the Meselson-Stahl experiment as crucial is now to show that both parts of Duhem's problem, the problem of untested auxiliaries and the problem of an exhaustive partition of theoretical alternatives (including the unconceived ones)[16], can be kept at bay. To this end he develops a holistic account of experimental mechanism that includes both a model of the mechanisms producing the phenomena and parts of the experimental system, among them "the characteristic manipulations and measurement devices used." (2009, 34) Baetu (2017) has criticised Weber's reconstruction and argued that the Meselson-Stahl experiment was inconclusive for the hypotheses considered. Instead, "it was part of a broader research project aiming to elucidate the mechanisms of DNA replication" (2017, 4.) – which ultimately led to the development of new experimental techniques. "Thus understood, the experiment extended over a decade or more. However, the crucial experiment account attributes all or most of the impact of the whole series of experiments to a single set of experimental results." (2017) In the same vein as Lakatos put it, an experiment becomes crucial only in historical reconstruction and within the context of a broader research program.

We believe that the Weber-Baetu debate rightly follows the trend diagnosed in Sect. 3.2. to view underdetermination and crucial experiments as an epistemic and factual problem rather than a logical and semantic one. In this way, the first aspect of Duhem's problem, the auxiliary hypotheses, becomes embedded into an experimental research program. The reference to turn to in the present context is of course Franklin's (2013) philosophical reconstruction of the history of modern particle physics. There Franklin distils a list of reliable strategies that in effect allow one to keep Duhem's problem at bay and address the related problem of theory-ladenness of large-scale particle experiments. Beauchemin's (2017) autopsy of measurements with the ATLAS detector[17] can be read as a continuation, into the days of LHC, of Franklin's (2013) history of the reliable experimental strategies and rules of data analysis that characterize contemporary elementary particle physics. We will take up some of these strategies in Section 7.2. and discuss how they permit us to consider the Higgs discovery as a crucial experiment. Let us however first assess Franklin's assessment of crucial experiments.

---

[15] Note that Duhem actually believed that experiments could be crucial. But this could not be inductively inferred from the data, but required the *bon sens* of the physicists. While *bon sens* might have been a useful notion in Duhem's days, it seems to us too vague for large-scale experiments in particle physics. At best one might take *bon sense* as an umbrella term for the detailed set of experimental strategies given by Franklin (2013).

[16] Cf. Stanford (2006) who shows that it is difficult to find cases where the underdetermination was not eventually resolved.

Franklin and Perovic (2015) compare two ground-breaking particle physics experiments. While they classify the discovery of parity violation as a crucial experiment, the discovery of CP-violation represented only a 'persuasive experiment'. "The difference lies in the length and complexity of the derivation linking the hypothesis to the experimental result, or to the number of auxiliary hypotheses required for the derivation." (2015, 85) Indeed, physicists had speculated about parity violation before, and the observed effect was maximal. CP-violation was completely unexpected, but most theoreticians quickly settled for it. Franklin and Perovic consider this acceptance as a "pragmatic solution of the Duhem-Quine problem." (2015, 84). In the case of the Stern-Gerlach experiment, as reconstructed by Franklin and Perovic, the diagnosis of cruciality underwent several changes. By discovering the space quantization [*Richtungsquantelung*] predicted by the Bohr-Sommerfeld quantum theory, it became a crucial watershed between classical and quantum physics, but not by confirming the latter theory. For what Stern and Gerlach actually measured was a new quantum phenomenon, electron spin, that was only postulated after the experiment. Thus, the experiment "was regarded as crucial at the time it was performed, but, in fact, wasn't. ... A new theory [quantum mechanics] was proposed and although the Stern-Gerlach result initially also posed problems for the new theory, after a modification of that new theory [the integration of spin], the result confirmed it. In a sense, it was crucial after all. It just took some time." (2015, 40-41)[18]

These examples also indicate that establishing experimental evidence and deciding whether an experiment is conclusive or even crucial, is largely a factual question and involves different time scales. Acquiring precision data sometimes represents a long-term process that involves previous experiments and is continued in the experiment itself. The actual discovery of a particle instead represents a precisely dated event; scientists decide after a detailed statistical analysis that the evidence is sufficient.

Using the Higgs discovery, in 7.2 we will argue that the diagnosis of Franklin and Perovic seems to us counterintuitive because it makes the characterization of an experiment as crucial or not depend on short-term development of scientific theorizing. In Section 7.2., we will provide a different characterization according to which all three examples mentioned qualify as crucial experiments.

## *4. The methods of this project*

---

[17] Note that Beauchemin's concept of theory-ladenness is wider than the one typically used in the philosophical literature, where theory-ladenness represents a problem for empirical science, not a feature that can be exploited by clever experimenters.

[18] In philosophical discussions about quantum mechanics, spin is considered as the quantum mechanical quantity par excellence and the Stern-Gerlach apparatus as its paradigmatic experiment. Notice that while Stern wanted to test quantum theory, Gerlach himself considered the experiment as part of a broader experimental research program.

Against the backdrop of the different experimental situations in 2011 and 2012, and the various solutions of the EWSB (including the Higgs mechanism) proposed by theoretical model builders, our project investigated the general attitudes and preferences of the LHC physicists by quantitative and qualitative empirical methods. In questionnaires and interviews LHC physicists were asked about their views of the status of particle physics, their anticipation of what the LHC will ultimately find, and the ways experimentalists and theorist interact.

Questionnaires were sent via e-mail to some 15000 physicists related to particle physics in August 2011 and September 2012. Each contained eight groups of questions, which were to be answered by either assigning a subjective probability for the correctness of a certain statement or by choosing an answer among various options. These were (i) the probability to find the SM Higgs particle (respectively confirm a minimal SM Higgs), (ii) the possible explanations of new physics found at LHC, (iii) the preference for certain BSM models independently of the LHC results, (iv) the criteria guiding the researcher's answers to this question, (v) the most critical flaws to the SM, (vi) the signatures in which LHC would most likely find new physics, (vii) general features of particle physics for whose understanding LHC will be most important, (viii) the interaction between experimentalists and theoreticians.

A large fraction of the questions within the above-mentioned groups were identical for the two periods, however, from experience with the first one, modifications were made for the second questionnaire. In the first questionnaire some answers could be ranked up to four times. This was considered less meaningful for the second questionnaire and modified. In 2012, a question was also added to address the Higgs boson candidate. The precise list of the questions can be found in appendix 2.

The lists of physicists to which the questionnaires were sent both in 2011 and 2012 were obtained from the INSPIRE data base (Dallmeier-Tiessen, S., Hecker, B., Holtkamp, A; 2016) maintained centrally at CERN. This data base is established by surveying journals, conferences, books, theses etc. in the pertinent fields and listing all authors. For the purpose of the questionnaire, authors in the categories 'hep-ph' (phenomenology), 'hep-th' (theory), 'hep-ex' (experiment) were contacted. In total this amounted to some 15000 authors. About half of the authors are theorists belonging to about the same amount to either the 'th' or 'ph' category, the other half experimentalists. Taking into account that some 8000 experimental physicists are directly involved in the LHC experiments, with an additional number of several thousand theorists, the list of physicists included probably almost all those who are actively working on LHC physics. Certainly, some physicists on the list were somewhat remote from LHC experiments or theory, e.g. mathematical theorists or accelerator physicists, but also some retired physicists or those who had left the field. It is difficult to assess, how large a fraction this was.

The anonymous replies were collected and statistically evaluated at Wuppertal. There were 1435, respectively 903 replies to the two questionnaires, which corresponds to a return rate of 10%, respectively 6% which is acceptable for empirical studies that are combined with interviews. Our goal had not been to obtain a truly representative sample in the sense of quantitative sociology. Still, there seems to be no strong bias in our replies: the regional distribution of respondents is consistent with the regional distribution of physicists working in LHC experiments, and also the fraction of theorists and experimentalists agrees with the fraction in the list. Yet, there are discrepancies as regards seniority: only few PhD students (<5% of the replies) have answered the questionnaires, whereas they amount to about a third in the LHC experiments. In the following, the replies were considered separately for experimentalists and theorists because this promised some interesting insights. In addition, the comparison of the replies before and after the discovery should indicate certain trends in the thinking of the LHC physicists.

In addition to the questionnaires, 9 (6) LHC physicists were interviewed around April 2011 (September 2012). Both groups included experimentalists from different LHC experiments and theorists. Furthermore, it was attempted to cover a wide range of interests and responsibilities within the LHC project. There is only a small overlap between the physicists in the two rounds; this was done deliberately in order to obtain a broader picture. The physicists interviewed and their respective roles at the time of the interviews are listed in Appendix 2. In the following discussion, no names will be assigned to the respective citations.

Each interview took about an hour. A few topics were addressed in every interview, for instance: in 2011, the prospects of a Higgs discovery, the perceived status of super-symmetry, and the chances to find new physics; in 2012, the impact of the Higgs discovery on the interviewee's research. On the other hand, the interviews were kept flexible to better understand the reasoning and preferences of each interviewee. This included, depending on the answers of the counterpart, also questions about the work environment, the methods of research, which outcome is expected at the LHC and why, and which outcome would be preferred on theoretical or pragmatic-aesthetic grounds.

## 5. The physicists' expectations in autumn 2011

At the beginning of our empirical study, the physical situation was characterized by an excellent performance of the LHC and its experiments. The year 2011 brought an unexpectedly large amount of data at the energy of 7 TeV. Based on this understanding, the LHC physicists performed simulation studies predicting that the whole range pertinent to the mass of a Higgs boson could be covered at the LHC within two years. On the other hand, although a broad range of searches for new effects had been performed by fall 2011, no sign for any of the many postulated extensions of the SM had been found. In particular, no indication for Supersymmetry was observed. Supersymmetry had been highly favoured by theorists, and it was predicted that its particles could be detected shortly after the LHC launch. Supersymmetry is the only BSM model that provides a strict constraint on the highest allowed mass of the Higgs boson (of about 130 GeV). The sensitivity of many of these searches for new physics reached the energy scale of about 1 TeV, at which the naturalness problem should have been resolved before the corrections become too high.

### 5.1 Outcome of questionnaires

In total 1435 physicists answered the questions, with the number of theorists (769) and experimentalists (696) being about the same. The number of replies to each of the questions differed only by a small amount. Assuming multinomial distributions and an outcome for an answer of 50%, these numbers imply a typical error margin on the answers of 1.5% for the total sample and 2% for each subgroup. The precise uncertainty depends on the number of answers given; the fewer there are, the larger is the relative uncertainty. Where relevant, the exact uncertainties will be provided.

### 5.1.1 The importance of the origin of mass

The high expectations that physicists had in the LHC to understand the mechanism of mass generation become most apparent in the replies to a question about the importance of LHC results for several key problems of current physics. Participants were asked whether they fully agreed, somewhat agreed, were undecided, somewhat disagreed, or fully disagreed with the statement: '*LHC results will be very important to understand ...*'. Close to 50% (48%/49% of the theorists/experimentalists) chose to 'fully agree', and close to 80% (77%/80%) at least 'somewhat agreed' on the importance of the LHC for the 'origin of mass'. Comparable results were obtained for two other topics from the SM, 'strong interactions' and 'flavour physics', while the outcomes for BSM physics were much lower. The as of then only undiscovered element of the SM was accordingly given the highest priority among all the potential features that could be found at the LHC.

### *5.1.2 Expectation on finding the Higgs Boson at LHC*

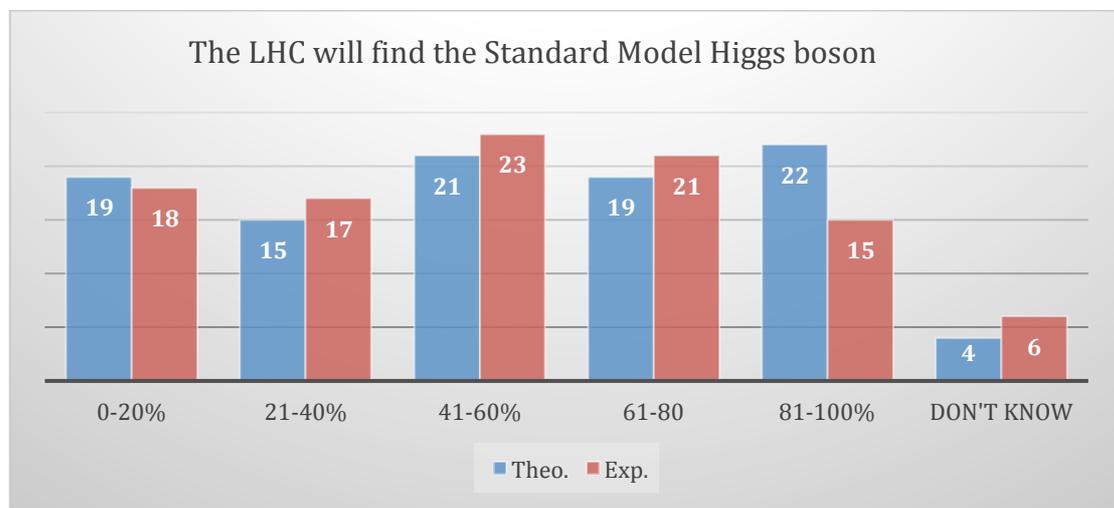

*Fig.1 Percentage of answers of theorists (blue) and experimentalists (red) assigning probabilities in intervals of 20% on the chance that the LHC will find a Standard Model Higgs Boson (Questionnaire of 2011)*

Given the importance of the origin of mass and the fact that the LHC was expected to provide the ultimate sensitivity for finding the SM Higgs boson the questionnaire asked physicists: '*What is your personal estimate of the probability [that] the LHC will find the Standard Model Higgs boson?*'. This (subjective) probability was to be given in terms of percentage intervals of 20%, which represented the respondent's current degree of belief. The replies did not reveal any strong tendency towards either discovery or non-discovery, but instead were rather uniformly distributed over all probability values (see Fig. 1). Some 35% (34% of theorists/35% of experimentalists) assigned a chance of at most 40% that the SM Higgs boson will be discovered, whereas only a few more (41%/36%) expected it to be found with 60% probability or more; the values for more than 80% probability were even lower (22%/15%). Thus, although simulation results showed that the LHC, in virtue of its foreseeable performance, had the potential to find the SM Higgs boson if it at all existed, a large fraction of LHC physicists assumed that it would not be found. These assessments were largely identical for experimentalists and theorists.

A second question addressed the '*personal estimate of the probability ... that the LHC will rule out the Standard Model Higgs boson*'. In this case 59%/46% of the theorists/experimentalists considered the probability low (i.e. smaller than 40%). On the other hand, only 19%/26% (uncertainty about 2.5%) estimated that the SM Higgs boson could be ruled out with high probability (i.e. larger than 60%). Low probability here means either that the SM Higgs boson will eventually be found or that a candidate is found whose properties cannot be measured precisely enough to rule out other interpretations. High probability instead means that the LHC will be able to definitively rule out the SM Higgs particle because there is no such particle or it will find one or more candidates that accomplish mass generation with properties different from the SM expectations. The responses showed that, in 2011, theorists were more sceptical about the LHC to rule out the SM Higgs boson than experimentalists.

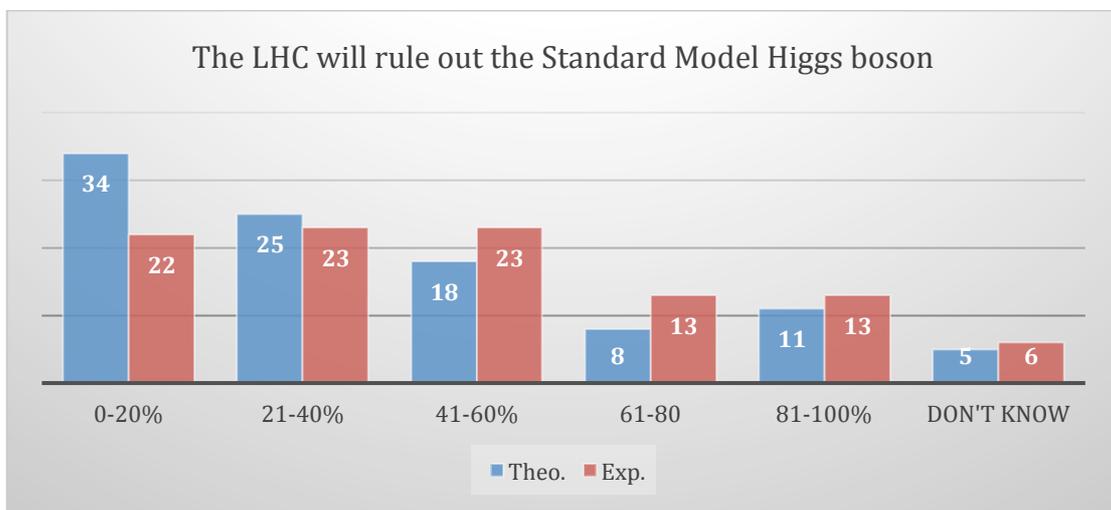

*Fig.2 Percentage of answers of theorists (blue) and experimentalists (red) assigning probabilities in intervals of 20% on the chance that the LHC will rule out a Standard Model Higgs Boson (Questionnaire of 2011)*

Although the two last questions are closely connected, there are subtle differences that lead to somewhat different replies. Firstly, the fraction of physicists that assigned an at least 60% probability to *find* the SM Higgs boson is smaller than the fraction of physicists who assigned an at least 60% chance that it will *not be ruled out*. Secondly, whereas the answers of theorists and experimentalists were rather consistent with the first question, a significantly larger portion of theorists than experimentalists considered it unlikely that the SM boson will be ruled out.

The first difference is probably related to the much stricter requirement to confirm not only the existence of a new particle, but to determine all of its properties to a precision that allows one, e.g., to distinguish it from alternative models of EWSB. Especially in the case of a more complicated Higgs group structure (as favoured by many physicists – see below) it will be more difficult to unambiguously identify the particle to be a SM Higgs boson than to rule it out. How-to interpret the differences between experimentalists and theorists is more difficult. In general, the replies – and the interviews below – indicate a greater reluctance of experimentalists to commit themselves to what their data will finally reveal.

This exemplifies the first lesson from the 2011 questionnaire. In contrast to public perception, interpreting the newly found particle as being the SM Higgs boson was not a simple yes/no alternative to be decided promptly. Physicists were largely prepared for a more complicated outcome that achieved all that the SM Higgs mechanism was designed for. Thus, finding a particle consistent with a SM Higgs would only be the first step in further investigating the properties of the new particle. The second conclusion from these two questions is that there existed a substantial scepticism among physicists as to the existence of a SM Higgs at this stage. This means that, although the LHC was expected to cover the whole allowed mass range for the SM Higgs particle, the LHC community was rather undecided if it exists. Taking both lessons together shows that there was no significant asymmetry in physicists' expectations between refuting and confirming the SM.

### 5.1.3 Expectations on various EWSB models

The questionnaire also addressed potential scenarios for 'new physics', i.e. a process or particle that is not part of the SM. Physicists were asked '*Assuming that the LHC finds new physics, which (if any) of the following models do you think has the best chance of explaining it'.* The physicists had two ranked choices; here we will typically just provide the first choice, the second gives fairly similar results. Several models, including those rather remote from EWSB, like string theory, extra spatial dimensions, or 4th generation models, were also considered. (cf. Appendix 2) In the following, we focus on the three most popular groups that were also those most closely related to EWSB. (Fig.3)

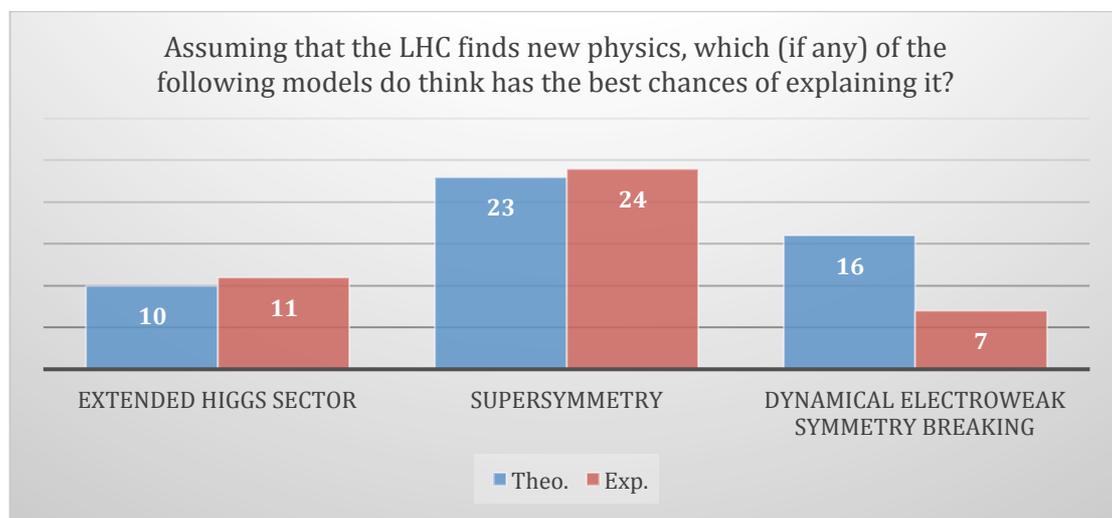

*Fig.3 Percentage of answers of theorists (blue) and experimentalists (red) on the most probable model that the LHC might find. Only answers with relation to the electroweak symmetry breaking are given, the remaining 51/58% refer to different models (Questionnaire of 2011)*

Fractions of 10/11% of theorists/experimentalists opted for an extended Higgs sector, i.e., more than one Higgs boson. About twice as many (23%/24%) voted for the favourite theory of Supersymmetry, which also requires an extended Higgs sector. Therefore, about one third of physicists were expecting new physics in extended Higgs sectors either without or within the context of an explicit model. Both of these answers assume Higgs bosons and expect them to be elementary as the SM Higgs. As mentioned above, one of the Higgs bosons in the extended sector might have properties very similar to the SM Higgs.

In addition, a sizeable fraction of theorists (16±1.3%) expected a dynamically generated electroweak symmetry breaking, leading to a composite scalar particle with several properties that are distinctively different from the SM Higgs boson. As discussed in Sect. 2.3., at least the historically first such model, Technicolour – a model that also contained the least additional assumptions –, had been strongly disfavoured by data. This might be the reason why only 7(±1)% of the experimentalists chose this option.

The follow-up question (Fig. 4) was '*which preference*' [physicists] have '*independently of the expectations regarding LHC results*', i.e., irrespective of the sensitivity of LHC itself. Whereas the replies alluding to an extended sector remained rather the same as to the previous question, dynamical EWSB was now even more favoured by theorists (19%), while the fraction of experimentalists preferring this alternative was unchanged at 7%.[19]

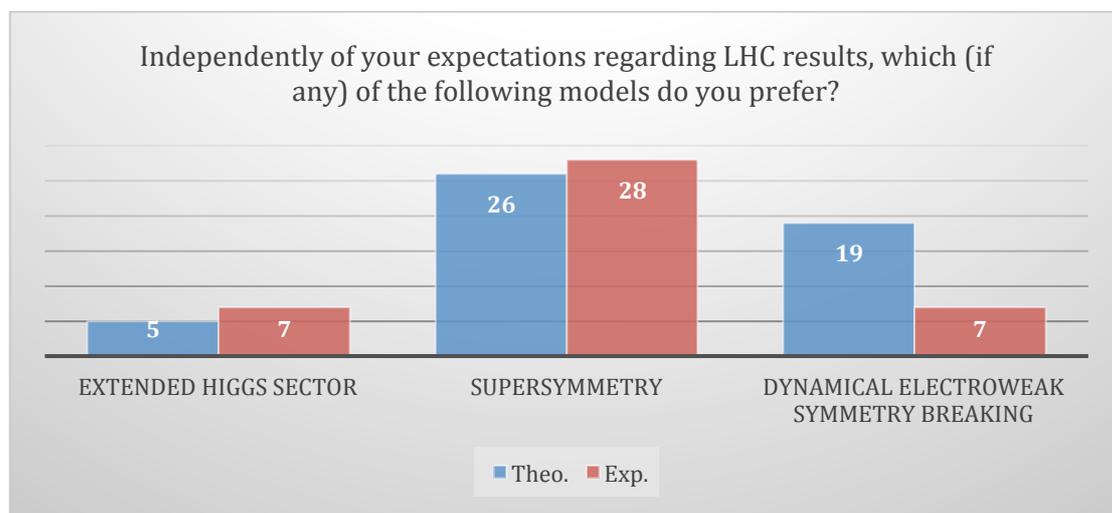

*Fig.4 Percentage of answers of theorists (blue) and experimentalists (red) on the preferred model for physics beyond the Standard Model. Only answers with relation to the electroweak symmetry breaking are given, the remaining 50/58% refer to different models (Questionnaire of 2011).*

The difference between the answers as to the importance of dynamical EWSB as a vision for the LHC data, on the one hand, and in a general perspective, on the other, shows that physicists' preferences are also guided by nonfactual and non-epistemic aspects. The perceived 'beauty' of a theoretical framework, or other pragmatic values of theory preference, weigh significantly relative to the chances of confirmation or disconfirmation by soon-to-be-available experimental data.

### *5.1.4 The importance of the naturalness criterion*

The question as to the preferred model was followed by the question '*which (if any) of the following criteria have guided you in answering the previous question?*'. Four ranked choices were allowed (Fig. 5). Considering only the first choice, the criterion that a model 'solves the naturalness problem' was preferred by 21%/17%. It is thus considered as important as the classical pragmatic, or rather aesthetic, criteria of 'elegance' (22%/18%) and 'simplicity' (16%/20%)[20]. More 'factual' criteria like the model 'will provide a better fit to the data', or 'makes specific predictions' or even 'has a candidate for dark matter' are much less considered (each one below 10% for both experimentalists and theorists).

---

[19] Although the present paper focuses on EWSB, it is worth noting that the largest percentages for new physics are obtained for the option 'None of those, but something totally unexpected' (28% for both theoreticians and experimentalists). The perspectives for new physics will be discussed in a separate paper that will also analyse some of the models neglected here.

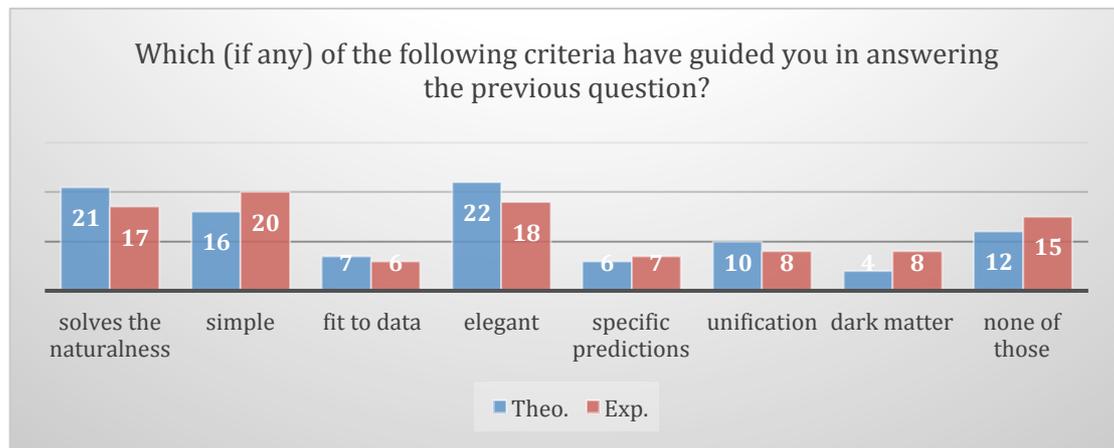

*Fig.5 Percentage of answers of theorists (blue) and experimentalists (red) on the criteria to choose the preferred model. (Questionnaire of 2011).*

Physicists were further asked '*what (if any) are the most critical flaws of the Standard Model*', and could make up to three unranked choices (Fig. 6). Indeed the problem of '*quadratic divergences in corrections to the Higgs mass*', causing the naturalness problem, was mentioned often by theorists (14%), while only by 10% of the experimentalists (statistical uncertainty of difference 2.2%). However, quadratic divergences are considered less of a flaw of the SM than its many parameters (18%/19%), the absence of gravity within the Standard Model framework (18%/21%), or that it does not include a Dark Matter candidate (17%/17%).

---

[20] It should be noted that physicists were not given any specific definition of these concepts; hence the replies were based on the intuition of the individual physicist. We do not consider this as too problematic for our purpose, not least because many philosophical authors who provide a definition - cf. Baker 2013 discussed in Section 3.2 – simultaneously emphasize that the terminology often is all over the place.

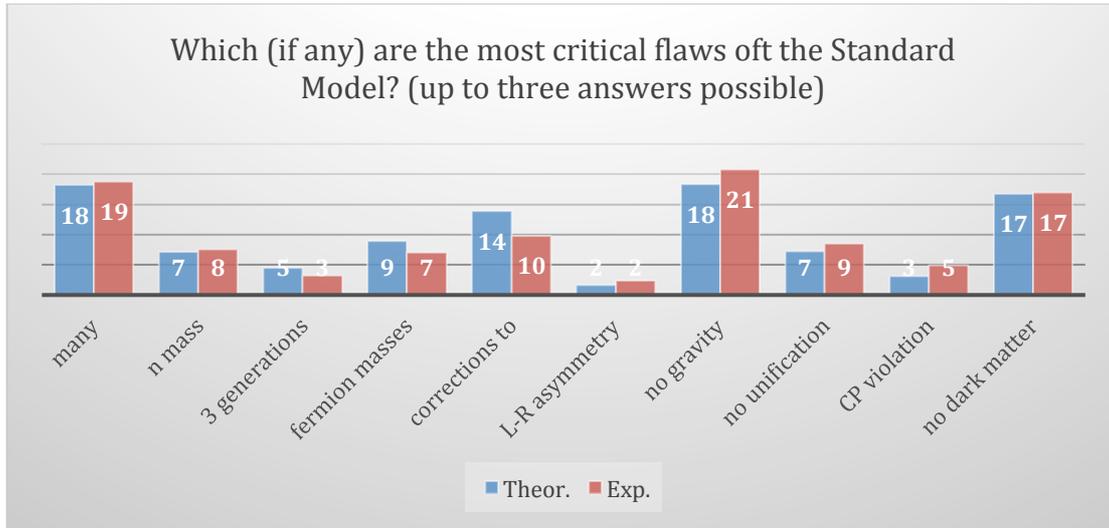

*Fig.6 Percentage of answers of theorists (blue) and experimentalists (red) on most critical flaw of the Standard Model (Questionnaire of 2011). The three answers were summed up and normalized.*

Both of these questions refer to SM properties that point beyond its limits. But they do so from a somewhat different perspective. The first (Fig. 5) had asked for the motivations of model preferences in Fig. 4, that is irrespective of the chances to soon find solutions at LHC. The second (Fig. 6) asked for flaws of the SM, irrespective of the existence of a credible model or strategy to resolve them. In both questions naturalness, respectively quadratic divergences, scored within the top group of the list and matched their counterparts, the pragmatic values of preference simplicity and elegance, and respectively (among the flaws), the many parameters of the SM.[21] However, the differences in the relative weights for other elements pointing BSM, e.g. for Dark Matter, are significant. To our mind, this has to do with the different perspectives of the questions. There are several proposals of physics BSM, however no universally agreed upon Dark Matter candidate. Nevertheless, it is a significant flaw of the SM, even though it may be only solved in the longer term. Naturalness, instead, is of immediate relevance to problems of model builders and guides expectations for BSM at the LHC. Experimentalists and theorists largely agree in this attitude.

*5.2 Responses in interviews*

The questionnaires were complemented by interviews with nine theorists and experimentalists[22]. Overall, their statements were consistent with the outcomes of the questionnaire. Yet they provide a deeper insight into the reasoning of elementary particle physicists at the time. In particular, they illustrate the rather diverse set of attitudes and the broad variety of expectations among the physicists. The following selected quotes are related to the mechanism of EWSB.

---

[21] The "many parameters" of the SM (Fig. 6) are traditionally seen as a principal lack of simplicity and motivate physicists to devise other models. In the same vein, the "quadratic corrections to the Higgs mass" (Fig. 6) amount to a peculiar technical feature in the renormalization scheme for the scalar Higgs boson that motivates models "solving the naturalness problem" (Fig. 5). In Section 5.2.3. we will provide some evidence from interviews that notwithstanding the philosophical distinctions discussed in Section 3.1., naturalness is largely treated in the same fashion throughout the community of particle physicists.

[22] For the list of names, see the table in Appendix 1.

### *5.2.1 Crucial, Long-Awaited, but Uncertain: Does the Higgs boson exist?*

The interviews were conducted at a time, when the allowed mass range for the SM Higgs boson was rapidly shrinking and the experiments were close to completely covering the remaining parameter space. No wonder that in all interviews the suspense whether the Higgs boson would be discovered or some alternative mechanism of EWSB would become visible, played a pivotal role. Here are two typical examples. One physicist stated that a discovery of the Higgs boson would amount '*to a revolution … We understand the mass, we understand a lot of things'.* Another one assessed the '*Higgs problem*' as a '*key question*'. The measurement of its mass should *,be a very important clue to what sort of theory maybe goes beyond it'.*

Although accordingly an experimental verdict, a crucial and long-awaited test for the SM, was in sight, opinions diverged on what its outcome would ultimately be. In this respect, the answers span a broad range. At one end of the spectrum, an experimentalist argued that in this situation one should '*press theorists*' to answer the question: '*if there is no such thing [the Higgs boson], then what?*'. Being a few femtobarns away from the final call about the Higgs boson, this represented the mood of some physicists that one had to move to a '*provocative question*'. The interviewee even identified a ,change of *mind-set'* because the - to date unsuccessful – experimental searches led to a general doubt whether the Higgs was a ,done deal''.

Other interviewees emphasized the personal and even emotional aspects of this increasingly pressing insecurity. E.g. '*I don't know, I don't know*': '*we have been waiting so long for this, …. there is … no concrete criterion to really judge whether [it] is more or less likely and emotionally, needless to say, I would like to see that as soon as possible, so I hope it's more likely that it comes out, but it is purely emotional because I do not want to wait another five years, but I have no idea.*' Another interviewee diagnosed a change of attitude. Previously colleagues might have argued that it '*is much more exciting to see nothing. But it was before LHC started. Now that things work so well, people are sure that the Higgs will be found in 2012, 2013, public opinion you know changed dramatically*'. It is '*psychologically very interesting*'. This strong desire to 'find something' also reflected the increasing gap between the enormous success of the SM predictions during the past 40 years and the fact that quite a few still unsolved questions remained.

Other experimental interviewees were rather optimistic to find the Higgs boson. One stated 'I *would be more surprised if they don't find the Standard Model Higgs because I think that the Standard Model with the Higgs mechanism at the moment is one of the best ways of explaining the masses of the gauge bosons and particle masses, so I would really like the Standard Model like Higgs*', or more pronounced '*I will be surprised if it is not found. I think it will be found at 120 GeV.*' (This was in accordance with the indirect and direct limits at that time.)

The attitude, according to which the Higgs is 'the best way' to explain masses had already been strengthened by indirect measurements disfavouring otherwise preferred alternative models. It also becomes apparent in replies from a theorist, who held that '*there is a lot of circumstantial evidence in favour of that [i.e. the Higgs boson], the case is not proved but that might well happen*'. This factual statement, however, is immediately put into perspective, when the same interviewee points to his preference '*I would find a lot of intellectual attraction in the dynamical symmetry breaking models*'.

### *5.2.2 Is the Higgs mechanism attractive?*

While the expectation as to the possible discovery of the Higgs boson was an issue of considerable suspense at that time, the motivations to expect or reject the Higgs mechanism, differ among the physicists. Already in the above statements, it became apparent that the mechanism, even if expected, was regarded with some reservations.

There were several values of theory preference in play. One theorist held that the Higgs mechanism is also disfavoured ‚*because of its minimal predictive power*'. Instead the alternative scenario of strong EWSB is intellectually favoured. '*Whereas the dream would be, in a dynamical theory of electroweak-symmetry breaking …, there is at least a conceivable possibility of making definite predictions, however they should turn out.*' This requires a '*somewhat bigger theory*'. The scepticism is also shared by experimentalists. One stated drastically, '*the Higgs is a totally ad hoc thing. ….. If the Higgs is not there, it will not surprise me.*' He argued that '*people had faith in it the way people have faith in God*'. Another experimentalist held, that even if the '*Higgs will be found, … the Higgs mechanism seems not elegant*' and there are '*very attractive theories without the Higgs*'.

In fact, only a few physicists interviewed emphasized the broader virtues of the Higgs boson beyond merely giving a solution for one specific problem. One theorist focussed on its role in the more encompassing theory of Supersymmetry, which as mentioned gives an upper bound of the Higgs mass. It would be '*very disappointing to find Higgs at a mass compatible with SM at high energies* [above 130 GeV]'. '*If the Higgs boson is light as suggested by the data [i.e. indirect measurements and left–over phase space masses], then presumably ….. super-symmetry is a prototype of such a weakly interacting extension of the standard Higgs mode*', i.e. the virtue of the Higgs mechanism is its accordance with a larger and generally favoured theory. On the other hand, the interviewee points out, that the favoured variant of supersymmetry, has been so tightly tested without finding anything. Therefore, one finds oneself in '*a weird situation* '.

Moreover, one theorist rejected the statement that the Higgs mechanism is complicated and ad hoc, but emphasised the virtue of introducing spontaneous symmetry breaking into particle physics in general. The Higgs discovery would then be seen as something new '*in the sense of new particles, …. but it is a break–through since you have the experimental test of spontaneous symmetry breaking*'. Moreover, this general idea would be '*not immediately thrown away*', if the Higgs boson would not be found.

### *5.2.3 The problem of naturalness*

Several interviews stressed the value of naturalness as a pragmatic guideline. Yet as regards its aesthetic aspects, opinions differed. One theorists became a proponent of Supersymmetry once it was shown to solve the naturalness problem; '*so for me the big, sort of change in my world view came when people pointed out that super-symmetrical particles could potentially control the quantum corrections and make the theory more manageable.*'. Another theorist, who was asked whether the naturalness or the hierarchy problem were serious, answered more cautiously: '*now to assess this, one goes back to these convictions somehow that the progress of science is always driven by an aesthetic judgement …. that goes beyond mechanical relations between formulas, equations and the need to see beyond. In other words, when you see some recurrence, when you see some "accident", it is natural for a scientist to consider the possibility that it is not an accident but there is something beyond and then this accident becomes natural. Now, this is not always correct, there are many accidents that we witness around that are not driven by the first principles but just accidents. So in that respect one can be wrong, but for the issue of naturalness, all of it, so called problems of the standard model, the picture is quite compelling.*' Only one experimentalist was explicitly asked about naturalness. Again, there was no strong commitment, but instead it was *'take[n] easy, it is a matter of taste'*. Note that the interviewees did not distinguish between naturalness, fine tuning of the quantum corrections, and quadratic divergences.

## 6. The physicists' response to the discovery of a Higgs candidate

A year after the first questionnaire was sent out and the interviews had been performed, a sufficient amount of data was collected by the LHC experiments to provide the desired sensitivity for a SM Higgs boson in almost the whole remaining mass range. Indeed, in July 2012, the observation of a new boson was announced. Since the signatures were consistent with the expectations, there was a very broad consensus that this particle was a very strong candidate for a SM Higgs. However, the properties known by then were few, and the precision of the measurements was still marginal[23]. The simultaneous searches for new effects BSM remained inconclusive, although the mass reach and sensitivity was extended.

Shortly after the announcement, a new questionnaire was sent to the same mailing list and a new round of interviews was performed. One of the main aims was to understand if, respectively how, the views and expectations of physicists had changed.

### 6.1 Outcome of the Questionnaire in 2012

The second questionnaire was sent out in September 2012. To a large part, the questions were identical to those of the first questionnaire. In this survey 903 physicists replied, among them 464 theorists and 439 experimentalists. The typical statistical uncertainty in the replies is therefore 1.7% for the whole sample and 2.4% for each of the subsamples. The relative uncertainty of the answers between the two questionnaires depends on whether the same or different physicists replied. In the former case the relative uncertainty would be very small, in the latter case some 2.3% (for the whole sample). Since the answers were given anonymously, there was no way to tell. Compared to the first round the possibility of ranked answers was restricted to a single choice only or to unranked options. The subjective degrees of belief in a statement (cf. 4.1.1.) were rephrased as 'fully agree', 'somewhat agree', etc.

### 6.1.1 What is the new particle?

Reacting to the discovery of the new particle, a set of questions was directed at its likely significance for the SM and beyond. The first statement to be evaluated was ‚*After the discovery of the new particle at 125 GeV, the LHC will confirm the minimal Higgs sector'*. The majority of both experimentalists and theorists (63%/63%) fully or somewhat agreed with this statement (see Fig. 7). Compared to the first questionnaire[24], this is, not surprisingly, a significant increase from the 41%/36%. Still, only 19%/23% 'fully agreed' that LHC will confirm the minimal Higgs sector of the SM.

---

[23] It was only half a year later, after more data became available and more studies had been made that the particle lost its status of being a candidate and was indeed considered a Higgs boson. This became apparent in a CERN press release (CERN press office, 2013) in which recent results were summarized.

[24] As a reminder, the exact wording of the first questionnaire was '*What is your personal estimate … that the LHC will find the Standard Model Higgs boson'*.

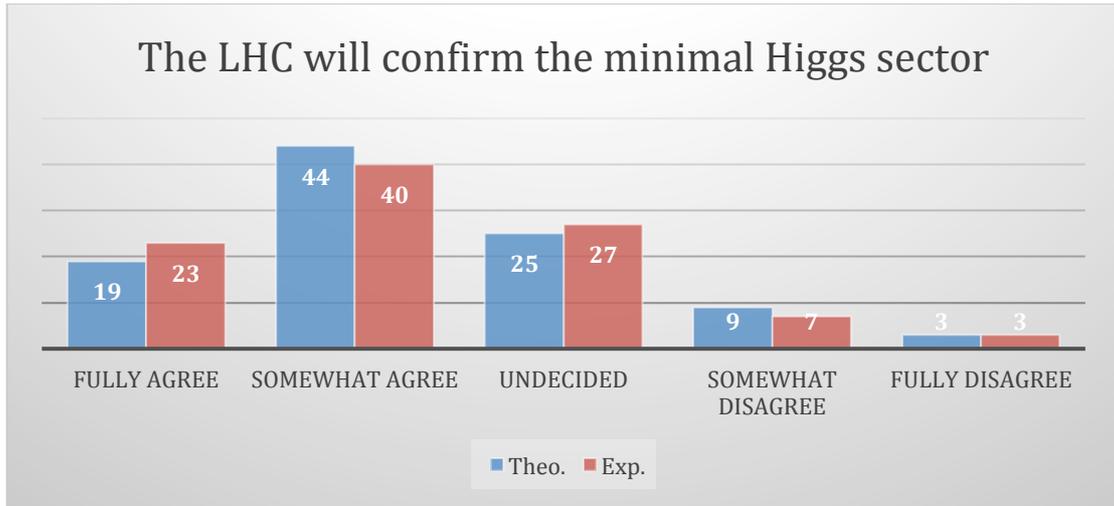

*Fig.7 Percentage of answers of theorists (blue) and experimentalists (red) assigning probabilities in intervals of 20% on the chance that the LHC will confirm a minimal Model Higgs Boson (Questionnaire of 2012)*

This question combined two aspects: whether the observed particle was indeed a Higgs boson and whether it would remain the only Higgs boson, i.e. the SM Higgs boson. This ambiguity could be somewhat resolved by asking a second question, to wit, whether the LHC will '*find a more complicated Higgs sector*'; for the new particle could be one of many Higgs bosons (Fig. 8). A sizeable fraction of 30%/31% (theorist/experimentalists) expected this to be the case. This is almost the remainder of those who fully or somewhat agreed with the first statement. However, almost half (46%/47%) of the responses were 'undecided', consistent for theorists and experimentalists. As in 2011 a more complicated Higgs sector appeared to be a very attractive option for many physicists. One may speculate about the reason for this rather neutral opinion. Certainly, the data were too scarce at the time of the questionnaire; moreover, the physicists may have been considering the probably limited precision of the LHC measurements.

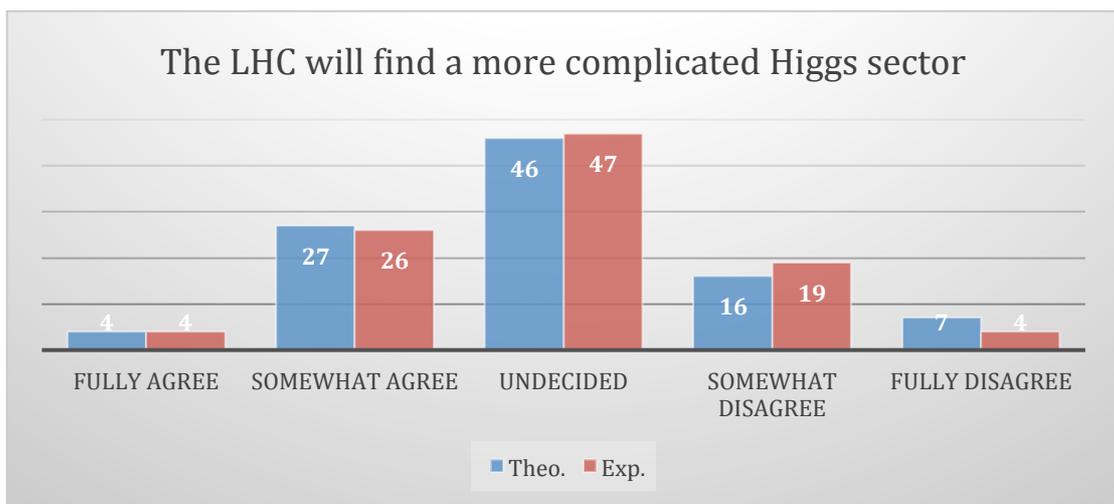

*Fig.8 Percentage of answers of theorists (blue) and experimentalists (red) assigning probabilities in intervals of 20% on the chance that the LHC will find a more complicated Higgs sector (Questionnaire of 2012)*

The third question asked if the LHC will '*find an alternative mechanism of EWSB*' (Fig. 9). It was only a minority of roughly 10% that agreed at least 'somewhat' with this statement. Full agreement was at the 1% level. Given the discovery of a Higgs candidate shortly before the questionnaire, such a result is not surprising. Even though the LHC data available at that stage were still marginal, the consistency with what is expected for a SM Higgs boson disfavoured a radically different mechanism already then. Although a third of the replies were undecided, the vast majority of physicists no longer expected any radically new physics to emerge in the sector of mass generation.

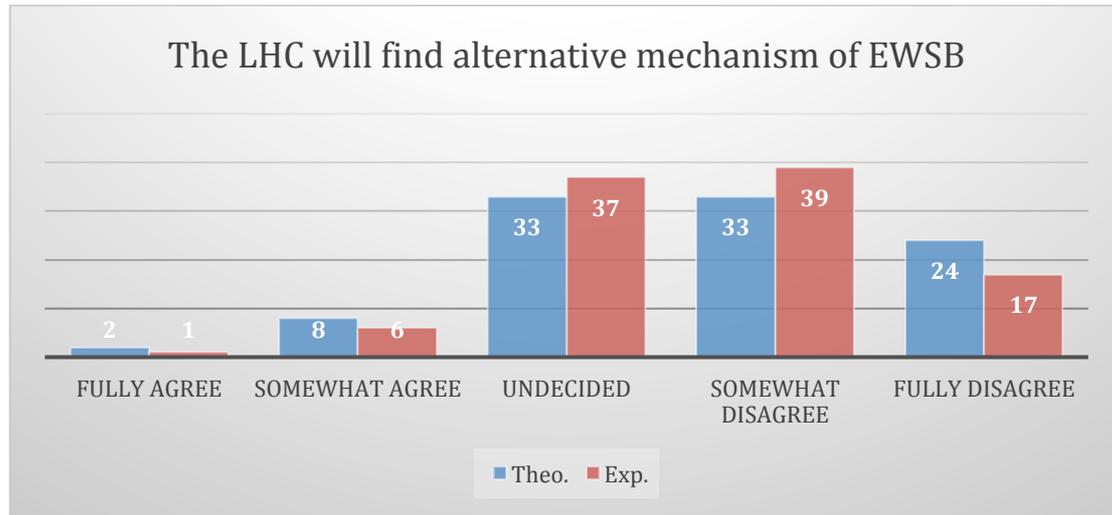

*Fig.9 Percentage of answers of theorists (blue) and experimentalists (red) assigning probabilities in intervals of 20% on the chance that the LHC will find an alternative mechanism of EWSB (Questionnaire of 2012)*

### 6.1.2 Which alternatives to the SM Higgs are still considered?

Whereas the new particle was largely considered to be a Higgs boson, its discovery initially did not preclude it to be, or involve, an element of new physics. Hence physicists were again asked: '*Assuming that the LHC finds new physics, which (if any) of the following models do you think has the best chance of explaining it?*'. However, while in 2011 two ranked choices were possible, only one was allowed in 2012 (Fig. 10). To compare the two surveys, only the first choice of 2011 is considered here.

Some 40% favoured extended Higgs sectors either with or without Supersymmetry, an increase from the about 34% in 2011. Going into more detail, the fraction of those who assumed Supersymmetry to explain new physics did not change with the discovery (25%/24% from 23%/24% as the first choice in 2011). Given that Supersymmetry was the only theory to predict such a light Higgs boson (cf. Sect. 2), its discovery could be seen to have strengthened the case for supersymmetry. On the other hand, none of the expected direct signals of Supersymmetry had been found, seemingly balancing the indirect support from the Higgs mass. General extended Higgs sectors have gained some ground in 2012 (increase from 10%/11%, as first choice in 2011, to 15%/14% in 2012).

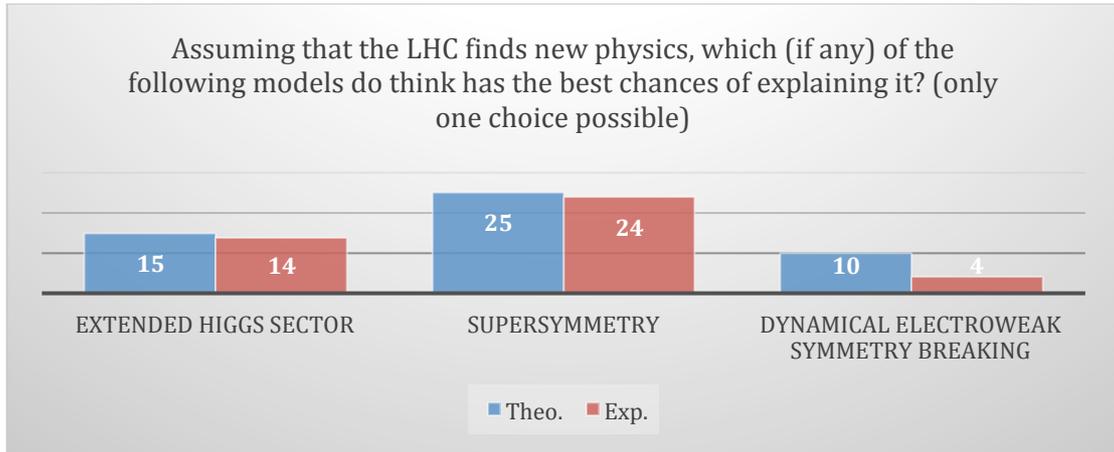

*Fig.10 Percentage of answers of theorists (blue) and experimentalists (red) on the most likely model for physics beyond the Standard Model. Only answers with relation to the electroweak symmetry breaking are given, the remaining 50/58% refer to different models (Questionnaire of 2012).*

In contrast, the fraction of physicists considering dynamical EWSB as the best chance was reduced to almost half between 2011 and 2012. Just 10%/4% of theorists/ experimentalists advocated for it after the observation of a Higgs candidate (previously 16%/7%). Even in spite of the limited parameter space of composite Higgs models, the number of proponents was still remarkable among theorists. The replies to the question ‚*Independently of your expectations regarding LHC results, which (if any) of the following models do you prefer?*' (Fig. 11), hardly changed. There was only a small decrease in the responses for dynamical electroweak symmetry breaking (14%/5% after 19%/7% in 2011). Thus, a significant minority among theorists prefers a solution of EWSB that is different from the Higgs mechanism, even though many do not believe this to be realised at LHC energies. This testifies the tenacity of theories that are attractive on internal grounds, that have strong pragmatic virtues, notwithstanding negative empirical results as long as there remain at least some options to adapt them to the data. (cf. 5.1.3.)

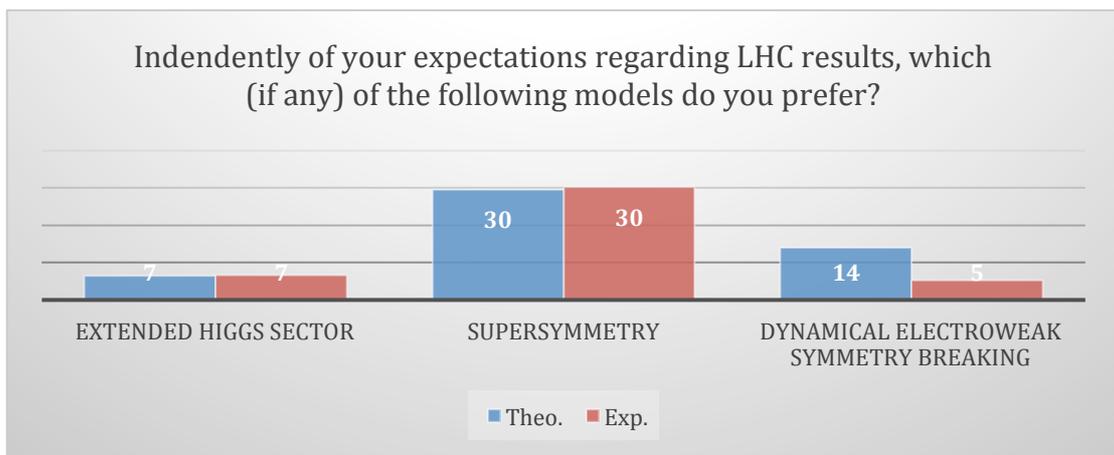

*Fig.11 Percentage of answers of theorists (blue) and experimentalists (red) on the preferred model for physics beyond the Standard Model. Only answers with relation to the electroweak symmetry breaking are given, the remaining 50/58% refer to different models (Questionnaire of 2012).*

### *6.1.3 How was the naturalness problem seen after the Higgs candidate?*

With the discovery of the Higgs candidate, i.e. the likely existence of an elementary scalar, it appeared that the naturalness problem had changed from a potential problem to an actual one. It could no longer resolve itself by the absence of a scalar from the set of fundamental particles. Furthermore, no BSM signal had been found to alleviate these concerns. We therefore tried to understand whether the physicists' assessment had changed after the Higgs discovery.

As in 2011, physicists were asked for the criteria, *'which have guided'* their selection of the preferred model, irrespective of the chances of confirming it at LHC. (Fig. 12) The attitude towards naturalness, however had hardly changed after the observation of a Higgs candidate. Naturalness was mentioned in 17%/18% of all answers, only mildly behind the criterion of ‚elegance' (20%/17%), but clearly ahead of simplicity (12%/10%).

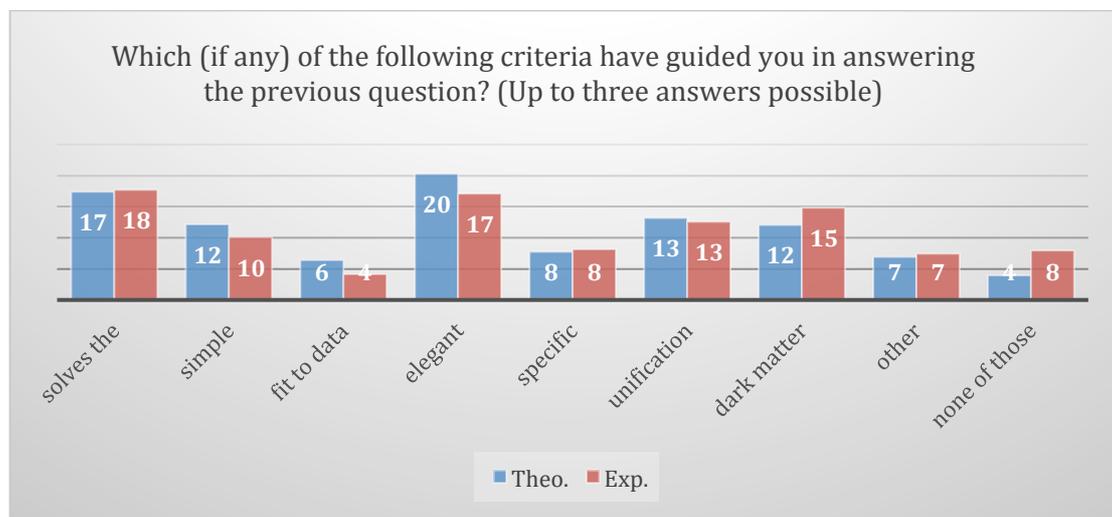

*Fig.12 Percentage of answers of theorists (blue) and experimentalists (red) on the criteria to choose the preferred model (Questionnaire of 2012). The three choices were added up.*

The other question pertinent to assessing the physicists' attitude towards naturalness was again the '*most critical flaw of the Standard Model'*. (Fig 13) As in the 2011 survey up to three choices could be given. As before '*quadratic divergences in corrections to Higgs mass'* remained at 15%/11% (compared to 14%/10% in 2011) as one of the three major flaws. Also most of the other assessments were fairly similar. A notable exception was that experimentalists now tended to consider the absence of a dark matter candidate' (21% after 17%) to become more critical. In both cases the replies from theorists did not change significantly. After the discovery of the Higgs boson, it seems, experimentalists shifted their interest to the next problem, which they thought to be in reach of the LHC, dark matter, even though there was no consensus about a suitable theoretical model, or whether such a dark matter candidate would be at all in the energy range of the LHC.

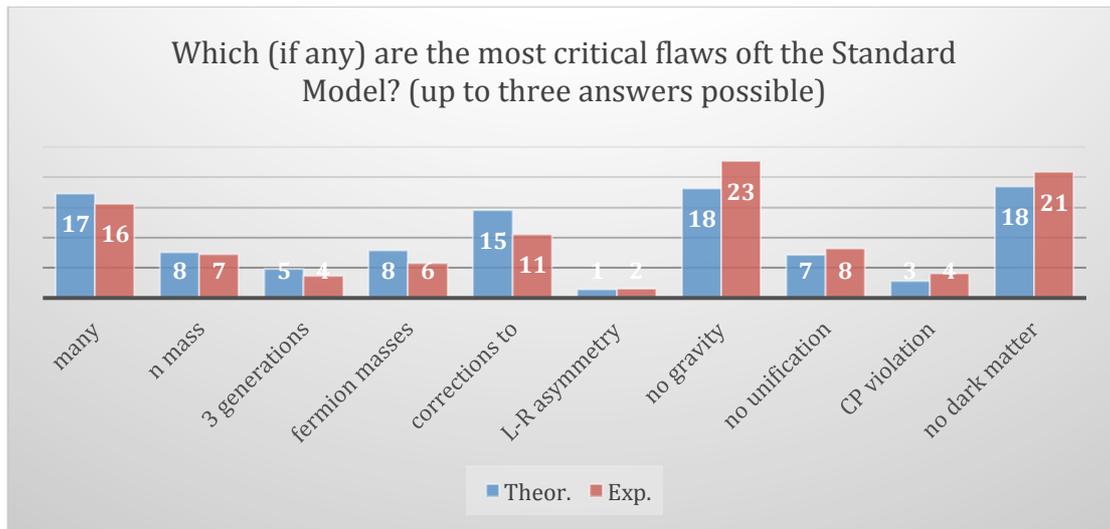

*Fig.13 Percentage of answers of theorists (blue) and experimentalists (red) on the criteria to choose the preferred model (Questionnaire of 2012). The three answers were added up and normalized. Note that the classifications are abbreviated the exact questions are listed in Appendix 2*

Both questions show that the perceived importance of the naturalness problem has not been affected by the change in the problem's specific status. A possible reason for this stability is that epistemic and pragmatic criteria, once adopted by the community, usually operate on a longer time scale. It takes a certain record of scientific successes to support them – as has been the case with the naturalness principle, at least in the eyes of the particle physics community.[25] That said, one may expect that after years of unsuccessful searches for BSM physics, naturalness would become much less attractive.

### 6.2 Outcome from interviews in the light of the discovery

As in 2011 the questionnaire was complemented by interviews, this time with five theorists and two experimentalists. Only one of them had already been interviewed in 2011; this time more emphasis was given to less senior physicists. The interviews took place in autumn 2012.

A large part of the interviewees characterized the Higgs discovery as an '*exciting*' event that would have decisive implications on future research. Overwhelmingly physicists cautioned to jump to immediate conclusions about the details of the SM Higgs boson. One interview partner expressed this very clearly '*well, we still do not know what we observed*'. But: '*It would have surprised me more if it would not be it [the Higgs boson]. This is in some sense paradoxial, since it is just something one got used to'.* Despite all caution there was agreement about the next steps.

### 6.2.1 From the observation to scrutiny

The focus, both experimentally and theoretically, was now to qualify this boson and look whether it was the SM Higgs boson or whether it had new physics in its wake. An experimentalist noted *'At this stage we observe a new particle … with properties*

---

[25] Guidice (2008) shows that there the record of the naturalness principle becomes mixed if one assumes a broader perspective; and one might even consider the fortunes of fine-tuning arguments more general. However, we are restricting our considerations to the understanding and role of naturalness within the community of elementary particle physicists.

*consistent with the Standard Model Higgs. This can change …. if we find something that does not fit into the Standard Model*'. But also theory is required to improve the precision of the predictions of Higgs properties '*If you don't see new physics directly, then maybe we see it through precision measurements, …. indirect tests.*'

In this sense, the new particle is seen as a potential harbinger of new physics, not the closure of a research program. 'Higgs physics' changed from searching for the Higgs boson towards measuring the properties of the new particle, assuming it to be 'largely' the Higgs boson, but also searching for deviations from its precisely predictable properties.

### *6.2.2 The implication on other models*

As discussed above, Supersymmetry predicted the Higgs boson to be lighter than 130 GeV. It would have been in deep troubles, had the Higgs been found at a larger mass. After unsuccessful searches for direct signs of Supersymmetry, the mere consistency of the measurements with its prediction, was taken as indirect support. One theorist says '*if it had been 140 GeV or 150 – that would have changed a lot, because then my SUSY-models would all be dead. And I would stop working on them.*'

Another impact of the Higgs discovery on BSM models is that, within the experimental uncertainties, it provides additional constraints. Indeed, in this perspective '*the discovery of the Higgs' is the main… result that influences our work*', a theorist explained; '*when we [work on] models we have to take… into account …  this particle and this changes… the situation definitely'.* Its existence and even the marginal precision of autumn 2012 constrains the allowed parameter space of BSM models, '*the determination of parameters of models or the testing of models I continue to do, now including the information from the Higgs. And that has changed something in the interpretation*'. E.g. to determine the allowed parameter space in supersymmetric models. '*What would be a 125 GeV Higgs [in super symmetry]? Of course, a SUSY Higgs, but what a parameter space would be compatible with this? It would be a very small one.*' In this sense the discovery of the Higgs candidate has severe implications for many models, at least in limiting significantly the allowed parameter range. This is of course an effect that becomes poignant on a larger timescale when it may eventually squeeze out certain models entirely. Such an effect was not yet visible in the changes of model preference between the 2011 and 2012 questionnaires.

### *6.2.3 Shedding doubt on the previous guiding principle of naturalness?*

As mentioned in 5.1.3, the naturalness problem in 2012 turned from fiction to reality. This can also be gathered from the interviews.  '*I would say that now that it is certain that there is a Higgs state at this mass, [the naturalness problem] is more alive than ever*'. And at least some continue to consider it an important question. Naturalness '*is still an important argument. I cannot see any reason, why should happen something like that, such [fine – tuning] just in a natural way. …. we put this fine-tuning by hand, … it cannot happen in nature.*'

Naturalness continues to be seen as an important guiding principle for the development of BSM models. ''*We need the guidelines. Because, it's not just the experiment, it's not just mathematics. It's something, which is between induction and deduction. … And you need … some guidelines. One guideline could be this naturalness, … which is a theoretical guideline. Or, 'minimality'*, one theorist argued, that is, for a model to have the minimal number of free parameters.

However, this was not the only reaction. The unnaturalness of the SM becomes more acute since none of the anticipated solutions in terms of New Physics has shown up in the energy range where it was expected, a fact that was considered highly disturbing. This dilemma leads to a growing discussion about the status of the naturalness problem: '*is this problem a real problem or just a fantasy of theoretical physicists? .... I've been trained to look at it as a serious problem*'. Yet another theorist raises doubts: '*what we thought was a main motivation to expect Supersymmetry at the LHC, this hierarchy problem or naturalness problem, .... I'm not so sure anymore whether this is actually something that leads us into the right direction.*' '*People have just accepted the fact that there is more and more fine-tuning now, because of the limits that become larger and larger. And it's not so clear to me whether it's still a good idea to consider that*'. Similarly another theorist argued, '*We can discuss whether... we have to accept fine-tuning. What could be behind a fine-tuning ... or if one wants a natural theory without fine-tuning. ... This is the main argument which ... I think, drove ... the theoretical community for the last twenty or thirty years. But it's not a solid argument'.*

To sum up, the discovery of the Higgs boson, together with the absence of any sign of new physics, has turned naturalness from a potential into a real theoretical problem. While before the discovery of the Higgs boson physicists might have expected that naturalness could be restored by a different mechanism of EWSB, LHC has now confirmed the unnatural Higgs mechanism, yet without finding evidence for a potential solution outside the EWSB sector. It is true, naturalness could still serve as a guideline for devising new models. But the absence of a cure for the naturalness problem of the SM has made some physicists wonder whether it is actually a deep problem or whether one should simply accept fine-tuning as a fact about nature and accept models that violate naturalness. Therefore, the naturalness problem was still considered to be important in the questionnaires, but the interviews showed that its previous importance was put into question. Throughout the interviews, there was no indication that physicists considered the naturalness problem as multifaceted or vague. Instead they interchangeably denoted it as hierarchy problem, fine-tuning, or quadratic divergences. In Section 7.4., we will discuss how this coherence in scientific practice squares with the differences in philosophical analysis.

## 7. Some philosophical lessons

In this Section we interpret the outcome of interviews and questionnaires in light of the questions mentioned in the Introduction.

### *7.1 Scepticism before and after a crucial test*

In retrospect, it might appear that the Higgs discovery had been largely expected. Our studies show that, in actual fact, the expectations of the LHC physicists were fairly diverse. At least shortly before a Higgs candidate was discovered, the community was basically split whether to expect the observation of a SM Higgs or not. (cf. 5.1.2) They all were aware that LHC had sufficient luminosity to accomplish such a crucial test. The reluctance to embrace the SM Higgs boson is in line with wide-spread criticism of the conceptual structure of the Higgs mechanism. None of the interviewees emphasised its theoretical elegance, some even considered it an ad-hoc argument. This reluctance is in contrast to especially Supersymmetry, which is frequently considered as too beautiful a theory that nature should not have chosen it – regardless whether it is realised in the LHC energy range. The proponents of the Higgs mechanism simply regarded it as the 'best' solution for mass generation that had come to the physicists' mind over the course of more than four decades. It did not contradict any measurement, was compatible with a wider theory, and had only relatively few parameters.

On the other hand, after a candidate with the 'right' mass and with properties consistent with the expectations had been observed, most LHC physicists-almost immediately embraced the notion that 'a' Higgs boson had been found – although they left it open whether it would be the only one. This overwhelming acceptance came, although the precision of the measurements still left quite some room for alternative solutions of EWSB. (cf. 6.1.1. & 6.2.1) In fact, one of the main research directions after the observation of the Higgs candidate, both experimentally and theoretically, became to scrutinize how large a parameter space for alternative solutions would be left. Whereas the vast majority did not consider solutions radically different from the SM, such as a composite scalar, as realistic, there was considerable hope to find deviations from the SM expectations that would give physicists a hint how to further investigate the complex landscape of BSM models. (cf. 6.1.2) When doing so, most data analyses after 2012 assumed the existence of a SM Higgs boson at 125 GeV, at least as the best approximation of the observed boson. This represented a significant discontinuity in the actual experimental strategy.

### *7.2. Was the Higgs Discovery a Crucial Experiment?*

In the interviews, the discovery of a Higgs boson was widely considered as extremely important for particle physics. This accords with the many statements in the literature during the last decades[26] and the significant material and intellectual resources that went into large experimental facilities to solve the problem of EWSB. This widely shared conviction among physicists that the problem of EWSB was at the crossroads of particle physics at large prompts the question as to whether the Higgs discovery represented a crucial experiment in a philosophical perspective? In Section 3.2 and 3.3, we have discussed underdetermination and its impact on the feasibility of crucial experiments and other ground-breaking experiments.

In this section, we argue that the underdetermination argument can be contained to such an extent that it does not play a role in actual scientific practice. On this basis we also argue that the Higgs discovery can indeed be considered as a crucial experiment. Even though the Higgs discovery – as shown in Section 7.1 and emphasised in many physics papers – was not a simple yes/no experiment, its cruciality, to our mind, rests upon the following characteristics:

a.  The Higgs boson was an essential and indispensable element of the SM. Moreover, the Higgs discovery was the final confirmation of a theoretical framework that had been developed over decades. Of course, the SM would also have broken down if, e.g. no $Z^0$ boson would have been found. But the Higgs, belongs to a sector of the SM that had not been seen before and is based on the additional concept of spontaneous symmetry breaking, for which no evidence had been observed before.

b.  The Higgs boson is fundamentally different from all particles found up to date because it is an elementary scalar. It is not the first scalar found, but e.g. the pion has been shown to be composed of two quarks and gluons.

---

[26] See, e.g. Ellis, J., Gaillard, M.K., and Nanopoulos D.V. (2012), Quigg (2007).

c. In view of the importance of the EWSB mechanism, a plethora of alternative models had been constructed. The discovery of the Higgs boson basically eliminated several families of these alternative models, and reshaped the direction of future research in a fundamental way.

Whereas a. and b. emphasise the crucial importance of a Higgs discovery for the SM and the general concept of elementary particle, characteristic c. widens the traditional philosophical understanding of crucial experiments which required that only one model survives. However, even if some alternatives may remain viable, the crucial experiment drastically reshapes the field. In the next paragraphs we argue that the crucial nature of an experiment as complex as LHC has to be judged against the backdrop of the historical development of the respective field.

Let us assess the acceptance of the observed particle being a Higgs in more detail. Immediately after the announcement at CERN in 2012 and the first measurements of its mass and decay modes, there was a flood of theoretical analyses, significantly reducing the possible parameter space of the many previously developed BSM models. Some models, among them 'higgsless models' and 'higgs-gauge unification', were strongly disfavoured and did not play a role in the subsequent discussions. Others, like 2HDM were more difficult to reject at this stage. As becomes apparent from the questionnaire, even only a few months after the discovery was announced, alternative models for EWSB were only expected by less than 10% of the respondents. (cf. 6.1.1.).

During the following year – with more data being analysed and additional properties being searched for, especially the spin – the notion of 'having found a Higgs boson' was adopted by the majority of the LHC physicists. The notion of 'a' Higgs boson leaves open the option of having a more complicated Higgs sector than the SM Higgs. In principle this can be resolved by higher experimental precision and additional searches, however, the precision will never be perfect in the future, such that small deviations from the SM Higgs cannot be ruled out with 100% certainty.

In this situation physicists are moving forward in their research accepting the SM Higgs boson to exist. This consensus was not based on logical inference, in the sense that physicists waited until all alternative solutions were definitively excluded.[27] It contained a certain dose of pragmatism in choosing promising research strategies. Such a 'pragmatic solution' to the Duhem-Neurath-Quine problem, as Franklin and Perovic (2015, 84) have aptly put it, does not preclude intensive future scrutiny of the signal both from an experimental and theoretical side. Just the opposite: this scrutiny leads to the emergence of a very significant new research direction. This persistent search for potential deviations is not, to our mind, in conflict with the Higgs discovery being crucial. To the contrary, such explorative searches that do not test models already proposed by theorists, can be seen as one way to address the problem of unconceived alternatives.

Let us discuss now more specifically how Duhemian underdetermination is openly addressed within the statistical data analysis. First, the discrimination of two (or several) hypotheses, framed in an identical theoretical environment, as the SM in particle physics, with the same (kind) of experiments is (almost) completely free of detailed theoretical considerations. For in such cases the well accepted and identical procedures are applied to either of the hypotheses. Duhemian underdetermination and theory-ladenness are shielded off by referring in the same way to an entire experimental set-up. This vastly reduces their sway among scientists.[28] Second, theory ladenness is significantly alleviated by using precision data from LHC and other experiments and the familiarity with experimental strategies, among them the rules of data analysis and the knowledge about background processes. The Higgs discovery, accordingly, was part and parcel of a longer tradition of accelerator experiments and the associated theoretical research programs.

---

[27] Wüthrich (2016) advocates a notion of diagnostic causal inference that partially dispenses with the explicit assumption of theories without sacrificing talk about causality in particle scattering.

One can formalize the physicists' handling of Duhem's problem as such. We denote the observed number of events of a certain signature[29] of scattering process by O, $T_1$ and $T_2$ are two theoretical hypotheses and $A_i$ a set of auxiliary hypotheses. The $P_i$ denote the predictions to measure a number of events given hypothesis $T_i$ and the auxiliary hypotheses. We then have[30] in a simplified notation omitting the uncertainties of $T_i$, $A_i$,

$(T_1 \& A_1 \& A_2 \& A_3 \& \ldots\ldots A_n)$  $|= P_1,$  $O |= P_1$

Duhem argues correctly that O can also be inferred from (e.g.)

$(T_2 \& A_1 \& A_2 \& A_3^* \& \ldots\ldots A_n)$  $|= P_1,$  $O |= P_1$

$A_3^*$ being an alternative auxiliary hypothesis. The two-pronged strategy by which particle physicists deal with the Duhem problem can be expressed as such. Using the identical auxiliary hypotheses for an experimental and theoretical environment allows one to test $T_1$ and $T_2$. Most importantly, each of the auxiliary hypotheses has been experimentally tested under multiple conditions. Theoretical assumptions are kept minimal and, if needed, they also have been tested extensively. Therefore $A_3^*$ can be excluded, leading to

$(T_2 \& A_1 \& A_2 \& A_3 \& \ldots\ldots A_n)$  $|= P_2,$  $O \sim|= P_2,$

which allows physicists eventually to discriminate between $T_1$ and $T_2$.

Let us be more specific about the auxiliary hypotheses. They include the known physics, rules of data analysis, criteria of statistical significance (cf. Beauchemin 2017). Some of these auxiliary hypotheses have been extensively tested before LHC was even built, others can be tested in–situ using the redundancy of LHC experiments.

In the actual practice of particle physics, the $T_i$ and $A_i$ are only known to some statistical and systematic uncertainty. This implies that also the predictions $P_i$ have uncertainties, i.e. $(P_i \pm \delta_i)$, where the $\delta_i$ are convolutions of the uncertainties of the individual $A_i$ and $T_i$. As a result, instead of a strict agreement or disagreement of the predictions with the observation O, only a finite likelihood $p_i(O| P_i)$ can be assigned taking into account the uncertainty of $P_i$. Duhem's problem therefore reappears in probabilistic terms. To keep underdetermination at bay, to resolve Duhem's problem in scientific practive, two additional conditions must be met. The first one is to confirm not only the correctness of the $A_i$, but also all their individual uncertainties. This is done simultaneously with the extensive tests of the $A_i$ (cf. Mättig 2019). Secondly, this probabilistic reasoning may in a strict (and naïve) manner be interpreted as impossibility to decide between $T_1$ and $T_2$, even if $p_2$ is ridiculously small, say $10^{-40}$. To 'exclude' such meaningless hypotheses $T_i$, the actual scientific practice defines conventions on the minimum magnitude of the $p_i$ to accept a hypothesis.

---

[28] Assuming the proper functioning of cables and switches is not to say that such errors do not occur. But it seems to us that those are not the kind of victories that advocates of a strong global notion of underdetermination (cf. 3.2.) would want to score. As Franklin (2013) shows, interesting failures of experiments are of a different kind.

[29] We will provide a more detailed analysis of the role of signatures in a subsequent paper that draws on the respective material from the questionnaire and the interviews.

[30] Note that |= does not amount to deductive entailment in the original Duhemian sense.

This shows that by embedding the actual experiment into a broader context and by distinguishing the different layers of theorizing and experimentation, crucial experiments are possible. Duhem's underdetermination can be addressed in scientific practice. This also agrees with Weber's intuition discussed in Sect. 3.3. that experiments are embedded into a broader experimental program and partially takes up Baetu's point that such a program may not only contain the purported crucial experiment. LHC has always tested other models and done exploratory searches alongside testing the SM.

Having dealt with Duhem's problem, let us discuss more generally what it takes to make an experiment in particle physics crucial. The Higgs discovery clearly was essential for the SM, but why was is not simply a 'persuasive experiment' in the sense of Franklin and Perovic? (cf. 3.3) After all, different mechanisms of EWSB are still being discussed even though – as we have seen in Sect. 6.1. – they are starting to draw less interest. Franklin and Perovic's distinction is based on the complexity of the inference |= and the number of auxiliary hypotheses required. But this seems to be problematic for large experiments like LHC that use many auxiliary hypotheses that are tested by different research groups. Taking into account also their analysis of the Stern-Gerlach experiment, it appears to us that the requirement of an immediate acceptance or refutation of competing models makes the notion of a crucial experiment subject to matters of short-term historical developments in the sense that that the status of the P violation experiment could have changed if some months later some alternative explanation had emerged that required further experiments to be excluded. Would the discovery of CP violation turn into a crucial experiment, if those alternatives had been devised before the experiment? Does the 'cruciality 'of an experiment depend on the number of models developed before and afterwards?

On the basis of these classical examples from particle physics, we are thus suggesting a notion of crucial experiment that is closer to experimental practice and less dependent on short-term developments in physical theory. This notion seems to us also in the spirit of Weber's (2009) discussion. But we are well aware that a substantive discussion of our proposal would require a broader set of examples, both positive and negative ones.

An experiment with a systematically and statistically significant outcome, is crucial in some field of science, if it is
a) seminal or decisive for the further development of this field,
and at least one of the following criteria are fulfilled.
b) It adds a new concept to the body of physics,
c) it implies a rejection of one or several theoretical solutions of a significant problem, or refutes an established concept.

The third criterion takes up characteristic b. in the above description of the Higgs discovery. In the examples discussed, P and CP violation fulfil criteria a) and c), the Stern-Gerlach experiment fulfilled criteria a), b), c) before the advent of quantum mechanics, fulfilled a), c) until spin was fully integrated into the theory, after which it fulfilled a) and b)[31]. The Higgs discovery fulfilled a) and b) by itself, but it also refuted alternative models, such that it also fulfils c).

---

[31] In contrast to Franklin and Perovic, we believe that the experiment's crucial character was present throughout. Initially, the experiment was proof of the phenomenon of space quantization

### 7.3. Principles and values of model choice

The questionnaires and the interviews have shown that in 2011 there was a large variety of epistemic and pragmatic values guiding the physicists' expectations about the Higgs searches, their concerns about the SM and their preferences for BSM physics. The discovery of the Higgs boson disfavoured certain models and strengthened the genuinely epistemic criteria. The example of dynamical symmetry breaking after the Higgs discovery showed that theoretical simplicity or other pragmatic criteria, e.g. fertility to calculate further particle masses, can motivate researchers even if the respective model is experimentally disfavoured. This corresponds to Kuhn's insight that the balancing of epistemic and pragmatic values is neither a perfect logical inference nor a matter of taste, but a fact in the history of scientific practice. Even though physicists are now largely accepting the SM as one of the most successful theories, they will keep looking for deviations that promise to be interesting.

Even taking into account tight experimental constraints, the 'expectation' to find dynamical EWSB was still significant among theorists before the observation of the Higgs candidate. It decreased after the observation, but remained remarkably high. Interestingly, the preference among theoreticians was, both in 2011 and 2012, always higher when asked 'independently of LHC'. (cf. 5.1.3. & 6.1.2.) Even though the chances for experimental evidence in the near future were low, dynamical EWSB remained a preferred solution for many. Notably, the preference of dynamical EWSB showed the largest difference in the surveys between experimentalists and theorists.

Let us evaluate possible consequences of applying the notion of values of preference to models instead of theories. Not all BSM models are as close to the status of a theory as the SM. Some of them are renormalizable and based on a sufficiently elaborated theoretical idea, such as Supersymmetry. But others are not; there is, for instance, considerable freedom in populating an extended Higgs sector. This does not prevent physicists from considering such models as worthy of pursuit, from choosing to investigate them, and indicate their motives for doing so. Finding evidence for such models would certainly have prompted theoretical investigations before physicists would commit themselves to the truth of that model in the same sense as one might commit oneself the existence of supersymmetry in nature. We do not see this as a weakness of our account, but as a consequence of the variegated model landscape.

Let us now look at our findings from the perspective of Douglas's classification of cognitive and pragmatic values. All models are physically consistent and are empirically adequate in the sense that they are not in conflict with the existing data. Thus, they fulfil the minimal criteria (i) and (ii), They also make specific predictions[32] and have a high predictive accuracy once basic parameter(s) are fixed. Furthermore, there is a consensus in the importance of the mechanism of EWSB. They are thus fulfilling the

---

predicted by the Bohr-Sommerfeld theory and a refutation of the classical views developed by Larmor. However, within quantum mechanics the concept of space quantization was replaced by a somewhat different notion of angular momentum, but this did not end the experiments crucial significance against pre-quantum theories (c). When quantum mechanics was fully established, including spin, there was so much spectroscopic evidence against the pre-quantum physics that the Stern-Gerlach experiment became less important in this respect (ending c)), but paradigmatic for the phenomenon of spin (fulfilling b with respect to this concept).

[32] The fractions of physicists choosing 'fit to data' and 'specific predictions' seems to be lower than one might expect but has to be seen on the background of these being not special to any model considered.

desiderata (iv). The questionnaire has also revealed a high score for genuinely pragmatic values that fall in Douglas's category (iii) and we have argued that they we able to balance the bleak empirical prospects for certain models as long as they were not ruled out in the sense of the minimal criterion (ii). Douglas admits that such a practice is legitimate as long as it is done with "the full acknowledgement that the theory is inadequate as it stands and that is must be corrected to meet the minimum requirements as quickly as possible." (2013, 802) She also admits that there are tensions in practice "between a well-supported theory (with group (iv) values supporting it) and an underdeveloped theory (with lots of group (iii) values and thus lots of potential)" (2013, 804). But these tensions are only pragmatic ones between conservatives and risk takers that do not endanger the separation between groups (iii) and (iv) because they "aim at different purposes" in the sense that "pragmatic criteria have no bearing on what should be thought of as our best supported scientific knowledge at the moment." (2013, 804)

It seems to us that when applying the values of preference to a complex model landscape, matters are a bit more complicated and Douglas' classification has to be taken with a grain of salt. Let us start with the claim that "groups (i) and (ii) … trump groups (iii) and (iv)" (2013, 804) and are clearly distinct as necessary conditions and desiderata. 'Internal consistency' (group (i)) in particle physics means that the theory is free of any infinities and can be extrapolated to energies $\Lambda \to \infty$. Strictly speaking, given the measurements of the Higgs and top mass this may even not be true for the SM itself, with the Higgs potential breaking down at high energy, although energies significantly beyond the Planck scale of $10^{19}$ GeV. This does not play any role for the $10^3$ GeV reachable at the LHC and the empirical adequacy of the theory within the experimental uncertainties. However, the same argument applies to the BSM models for EWSB. For all models, some energy scale $\Lambda$ is introduced at which some theory should exist – with properties that are vaguely known - and where it is assumed to be fully renormalizable. This $\Lambda$ is, in general, far beyond the reach of the LHC and the models predict just some 'low energy' (i.e. in the LHC range) phenomena, where the details of the full high energy theory do not play a role. As seen in the questionnaires and the interviews, this is not of concern for physicists. What is more important to them are solutions of problems like dark matter, unification, and naturalness (see next section), but also pragmatic criteria like simplicity and elegance.

In all these models, empirical adequacy is guaranteed by constructing them so as to encompass the SM that had a very high degree of experimental confirmation, except for the Higgs sector. While accordingly, empirical adequacy is accepted as a preeminent value, in the practice of physicists, internal consistency plays only a role if it prevents clear predictions. This means that group (iv), at least temporarily, can trump group 1. It is true, physicists are fully aware of this fact, however to reach $\Lambda_{BSM}$ may take decades, if not centuries. To evade this, one either has to redefine the meaning of physical consistency or retrigger the epistemic values. Wells (2012) considers mathematical consistency (his term for internal consistency) to be as preeminent as empirical consistency (i.e. adequacy), while not denying that this is not the general attitude among physicists. To implement this preference, he advocates effective field theories (EFT) that accomplish internal consistency by adding an infinite series of additional terms made up of all fields of the model, implying also an infinite set of free parameters. Whereas this is in principle correct it is hardly a practice followed widely in particle physics. When EFTs are considered in the actual practice, only a limited set of terms is used destroying the mathematical consistency but making the theory tractable.

### 7.4 The Guiding Principle 'Naturalness'

Our questionnaire has shown that naturalness is indeed on a par with the traditional pragmatic values of 'elegance' and 'simplicity' – as for the guiding principles of model preference – and on a par with 'too many independent parameters', the 'missing dark matter candidate' and the 'non-inclusion of gravity' – as for the most critical flaws of the SM. While the many independent parameters render the SM not simple – for decades elementary particle physicists have been looking for a simple unifying theory – the two other flaws concern empirical facts that cannot be accommodated by the SM.
From the interviews (cf. 6.2.3) we have concluded that naturalness is considered as sufficiently well entrenched within the community to be considered as a coherent guiding principle for scientific practice. But it operates both in an epistemic and a pragmatic mode.

Renormalization is the way to guarantee the finiteness of the theory, that is, its theoretical consistency. In principle the huge 'unnatural' renormalization corrections for the elementary Higgs boson do not make the theory inconsistent in a strictly formal sense. In practice, however, most physicists find this inacceptable - they do not accept too much fine-tuning of SM parameters – and try to find a remedy by supplementing the SM. This renders naturalness an epistemic value, in Kuhn's and Douglas's terminology. However, in our understanding, naturalness also acts a pragmatic value. It is an operationally relatively easy-to-apply quantitative criterion, at least once it is specified how much fine-tuning is allowed, and it constrains models; e.g. it suggests new particles with top flavour to compensate the main culprit for 'unnaturalness'. This may also be the reason why naturalness is maintained as an important criterion to devise BSM models. Such a double-track value of preference complicates the grouping of values for model preference. There is, to our mind, no clear separation of this complex criterion into more elementary epistemic and pragmatic values. For despite its complexity, naturalness is coherently applied as guiding principle by the physics community. It seems to us that this diagnosis does not contradict the philosophical differentiations advocated by Wells and Williams (cf. 3.3).

With the Higgs discovery 'naturalness' turned from a potential problem of the SM into a real one. The positive empirical finding was not accompanied by an observation of new particles in the TeV range that could resolve the problem in close temporal proximity to the Higgs discovery. One might expect that after confirming the cause of the naturalness problem, its solution should have been considered as more urgent. Such a trend was not visible in the questionnaire; its high ranking as guiding principle or most critical flaw of the SM did not change. The interviews revealed a more differentiated picture[33]. Some physicists still regard the naturalness problem as a nuisance, but contemplate that it might be an accidental feature of particle physics instead of a solid theoretical argument. In a sense, physicists are becoming prepared to live with it.[34]

The resilience of the naturalness problem may result from the fact that there exists no clear threshold when a theory becomes 'unnatural'. At least before the first results of the LHC folklore had it that fine-tuning requires new physics at an energy scale of 1 TeV. Yet there is no prohibitive argument against changing this to 5 or 10 (or more) TeV, even if this increases the amount of fine-tuning. Therefore, with some adjustments, the naturalness problem can still be maintained for the forthcoming LHC data; moreover, also the parameter space for new physics at the 1 TeV scale has not been completely covered by previous searches. Thus, many physicists defer the final word on the importance of naturalness to the higher energies and intensities that the LHC is about to enter. It remains to be seen whether BSM physics is found that indeed can cure the naturalness problem, If not, one may wonder how far the scale of new physics can be stretched or whether naturalness will eventually be abandoned. This again shows that it is much more specific than the usual pragmatic values of model preference.

---

[33] This agrees with several articles by physicists reconsidering the Naturalness problem, e.g. Guidice (2013), Dine (2015).
[34] Cf. Friederich, Harlander & Karaca (2014, Sect. 7).

*8. Conclusion*

Let us sum up the main results of our paper. First, the discovery of the Higgs boson and the confirmation of the SM were less expected than is often assumed. With the growing evidence that the newly discovered particle has properties consistent with the SM expectations, most physicists accepted it to be a Higgs, and at least tentatively, a SM Higgs boson. This does not contradict the fact that searches for possible deviations from the SM will be ongoing for a long time. Second, the Higgs discovery represented a crucial experiment for the SM if one interprets the notion in a sense that is appropriate for modern experiments. An experiment as complex as LHC cannot be properly understood without its embedding into a tradition of previous precision experiment and the tradition of reliable and established experimental strategies. These are crucial for keeping underdetermination at bay. Third, our case study suggests that criteria of theory choice be understood as epistemic and pragmatic values that have to be weighed in in factual practice. The Higgs discovery led to a certain shift from pragmatic to epistemic values as regards the mechanisms of EWSB. Complex criteria, such as naturalness, combine different values without becoming inconsistent or inapplicable by the scientific community.

**Appendix 1 List of interview partners**

**March & April 2011**

| Prof. V.Sharma | UC San Diego, USA | experimentalist | Convenor of Higgs group in CMS expt. | Male |
| --- | --- | --- | --- | --- |
| Prof. F.Gianotti | CERN | experimentalist | Spokeswoman of ATLAS expt. | Female |
| Prof. G.Tonelli | U of Pisa (Italy) | experimentalist | Spokesman of CMS expt. | Male |
| Prof. A.Golutvin | IC London (UK) | experimentalist | Spokesman of LHCb expt. | Male |
| Dr. J.Boyd | CERN | experimentalist | Coordinator data preparation in ATLAS expt. | Male |
| Prof. J.Ellis | CERN | theorist |  | Male |
| Prof. C.Quigg | Fermilab (USA) | theorist |  | Male |
| Prof. M.Mangano | CERN | theorist |  | Male |
| Dr. M.Mihalla | KIT (Germany) | theorist |  | Female |

**Fall 2012**

| Prof. M.Krämer | RWTH Aachen | theorist | | Male |
| --- | --- | --- | --- | --- |
| Prof. L.Feld | RWTH Aachen | experimentalist | | Male |
| Dr. L. Di Luzio | KIT (Germany) | theorist | | Male |
| Dr. F. Domingo | KIT (Germany) | theorist | | Male |
| Prof. C.Issever | U of Oxford (UK) | experimentalist | Convenor exotics ATLAS expt. | Female |
| Dr. M.Mihalla | KIT (Germany) | theorist | | Female |

**Appendix 2: List of questions in questionnaires**

**In 2011**

1. **What is your personal estimate of the probability of the following scenarios? The LHC will...**
   a. find the Standard Model Higgs boson
   b. rule out the Standard Model Higgs boson
   c. find indisputable evidence of new physics
   The probabilities to be assigned were in 20% intervals, i.e. 0-20, 20-40%, .......

2. **Assuming that the LHC finds new physics, which (if any) of the following models do you think has the best chances of explaining it**
   a. extended Higgs sector
   b. supersymmetry
   c. extra-dimensions
   d. dynamical electroweak symmetry breaking
   e. 4th generation
   f. extended gauge symmetry (Z', Little Higgs)
   g. string theory
   h. other
   i. None of those, but something totally unexpected
   j. I don't know
   The questionnaire asked for two ranked choices.

3. **Independently of your expectations regarding LHC results, which (if any) of the following models do you prefer?**
   a. extended Higgs sector
   b. supersymmetry
   c. extra-dimensions
   d. dynamical electroweak symmetry breaking
   e. 4th generation
   f. extended gauge symmetry (Z', Little Higgs)
   g. string theory
   h. other
   i. I don't know
   The questionnaire asked for two ranked choices.

4. **Which (if any) of the following criteria have guided you in answering the previous question? (Four ranked answers were possible.)**
   a. The model solves naturalness/hierarchy problem
   b. The model is simple
   c. The model will provide a better fit to the data
   d. The model is elegant
   e. The model makes very specific predictions
   f. The model allows the unification of forces
   g. The model has a candidate for dark matter
   h. other
   i. none of the above

5. **Which (if any) are the most critical flaws of the Standard Model? (up to three answers possible)**
   a. too many independent parameters
   b. small but nonzero neutrino masses

c. replication of fermion families
   d. different magnitude of scales of fermion masses
   e. quadratic divergences in corrections to Higgs mass
   f. left-right asymmetry
   g. gravity is not included
   h. no unification of strong and electroweak forces
   i. CP violation
   j. No dark matter candidate

6. **In which of the following signatures (if any) do you think the LHC will most likely find new physics?**
   a. signatures with bottom quarks
   b. signatures with top quarks
   c. signatures with tau leptons
   d. signatures with missing energy
   e. signatures with multi – jet topologies
   f. signatures with multi – lepton topologies
   g. soft events
   h. other
   i. I don't know
   Two ranked choices were asked for

7. **How much do you agree with the following statements? LHC results will be very important to understand…**
   a. strong interactions
   b. flavour physics
   c. origin of mass
   d. quantum gravitational effects
   e. dark matter
   f. dark energy
   g. cosmology of the early universe
   The answers should be given for each field in terms of
   'fully agree', 'somewhat agree', 'undecided', 'somewhat disagree', 'fully disagree'

8. **How much do you agree with the following statements?**
   a. There is plenty of dialogue between theoretical and experimental physicists on LHC physics
   b. Theorists are fully prepared to tackle future new data from LHC
   c. Theorists are making helpful suggestions on how to collect and analyse LHC data
   d. Experimental physicists are sufficiently taking into account suggestions from theorists
   e. Experimental physicists are presenting their results in the most helpful way for theorists
   The answers should be given for each field in terms of 'fully agree', 'somewhat agree', 'undecided', 'somewhat disagree', 'fully disagree'

**In 2012**

1. **How much do you agree with the following statements? After the discovery of the new particle at 125 GeV, the LHC will…**
   a. confirm the minimal Higgs sector
   b. find a more complicated Higgs sector
   c. find an alternative mechanism for EWSB

d. find indisputable evidence of new physics

   The answers should be given for each field in terms of 'fully agree', 'somewhat agree', 'undecided', 'somewhat disagree', 'fully disagree'

2. **Assuming that the LHC finds new physics, which (if any) of the following models do you think has the best chances of explaining it**
   a. extended Higgs sector
   b. supersymmetry
   c. extra-dimensions
   d. dynamical electroweak symmetry breaking
   e. 4th generation
   f. extended gauge symmetry (Z', Little Higgs)
   g. string theory
   h. other
   i. None of those, but something totally unexpected
   j. I don't know

   Only one choice was possible

3. **Independently of your expectations regarding LHC results, which (if any) of the following models do you prefer?**
   a. extended Higgs sector
   b. supersymmetry
   c. extra-dimensions
   d. dynamical electroweak symmetry breaking
   e. 4th generation
   f. extended gauge symmetry (Z', Little Higgs)
   g. string theory
   h. other
   i. I don't know

   Only one choice possible

4. **Which (if any) of the following criteria have guided you in answering the previous question?**
   a. The model solves naturalness/hierarchy problem
   b. The model is simple
   c. The model will provide a better fit to the data
   d. The model is elegant
   e. The model makes very specific predictions
   f. The model allows the unification of forces
   g. The model has a candidate for dark matter
   h. other
   i. none of the above

   Up to three answers were asked for

5. **Which (if any) are the most critical flaws of the Standard Model? (up to three answers possible)**
   a. too many independent parameters
   b. small but nonzero neutrino masses
   c. replication of fermion families
   d. different magnitude of scales of fermion masses
   e. quadratic divergencies in corrections to Higgs mass
   f. left-right asymmetry
   g. gravity is not included
   h. no unification of strong and electroweak forces

i. CP violation
   j. No dark matter candidate

6. **In which of the following signatures (if any) do you think the LHC will most likely find new physics?**
   a. signatures with bottom quarks
   b. signatures with top quarks
   c. signatures with tau leptons
   d. signatures with missing energy
   e. signatures with multi – jet topologies
   f. signatures with multi – lepton topologies
   g. soft events
   h. other
   i. I don't know
   Two ranked choices were asked for

7. **How much do you agree with the following statements? LHC results will be very important to understand…**
   a. strong interactions
   b. flavour physics
   c. origin of mass
   d. quantum gravitational effects
   e. dark matter
   f. dark energy
   g. cosmology of the early universe
   The answers should be given for each field in terms of 'fully agree', 'somewhat agree', 'undecided', 'somewhat disagree', 'fully disagree'

8. **How much do you agree with the following statements?**
   a. There is plenty of dialogue between theoretical and experimental physicists on LHC physics
   b. Theorists are fully prepared to tackle future new data from LHC
   c. Theorists are making helpful suggestions on how to collect and analyse LHC data
   d. Experimental physicists are sufficiently taking into account suggestions from theorists
   e. Experimental physicists are presenting their results in the most helpful way for theorists
   The answers should be given for each field in terms of 'fully agree', 'somewhat agree', 'undecided', 'somewhat disagree', 'fully disagree'

# References:

Aad, G. *et al*. (2012). Observation of a New Particle in the Search for the Standard Model Higgs Boson with the ATLAS Detector at the LHC. Physics Letters B, 716, 1-29. *arXiv:1207.7214v1 [hep-ex]*,

Baetu, T. (2017). On the Possibility of Crucial Experiments in Biology. The British Journal for the Philosophy of Science, published online.

Chatrchyan, S. *et al*. (2012). Observation of a New Boson at a Mass of 125 GeV with the CMS Experiment at the LHC. Physics Letters B, 716, 30-61.

Baker, A. (2013) "Simplicity", *The Stanford Encyclopedia of Philosophy* (Fall 2013 Edition), Edward N. Zalta (ed.), URL = <http://plato.stanford.edu/archives/fall2013/entries/simplicity/>.

Beauchemin, P.-H. (2017). Autopsy of measurements with the ATLAS detector at the LHC, *Synthese* 194, 275-312.

Borrelli, A. (2015). Between logos and mythos: narratives of "naturalness" in today's particle physics community. In "Narrated Communities: Narrated Realities", H. Blume & C. Leitgeb (Eds.)

Borrelli, A. (2016): Was Sie schon immer über das CERN wissen wollten, aber bisher nicht zu fragen wagten – eine philosophische und soziologische Perspektive, in H. Satz, P. Blanchard, C. Kommer (eds.), Großforschung in neuen Dimensionen Denker unserer Zeit über die aktuelle Elementarteilchenphysik am CERN, Berlin-Heidelberg: Springer, pp. 119-149.

Borrelli, A. & Stöltzner, M. (2013) Model Landscapes in the Higgs Sector. In *EPSA11 Perspectives and Foundational Problems in Philosophy of Science*, V. Karakostas & D. Dieks (eds.), The European Philosophy of Science Association Proceedings 2, Dordrecht: Springer, 241-252.

Carnap, R. (1950). Empiricism, Semantics, Ontology. *Revue Internationale de Philosophie,* 4, 20-40.

CERN press office (March 14, 2013) New results indicate that particle discovered at CERN is a Higgs boson; https://press.web.cern.ch/press-releases/2013/03/new-results-indicate-particle-discovered-cern-higgs-boson

Csaki, C. and Tanedo, P. (2016). Beyond the Standard Model Lectures at the 2013 European School of High Energy Physics, arXiv:1602.04228v1 [hep-ph] (2016)

Dallmeier-Tiessen, S. (2016), Hecker, B., Holtkamp, A. et al for the INSPIRE collaboration, How partnership accelerates Open Science: High-Energy-Physics and INSPIRE, a case study of a complex repository ecosystem, CERN-OPEN-2016-008.

Dine, M, (2015). Naturalness Under Stress, Ann.Rev.Nucl.Part.Sci. 65: 43-62, arXiv:1501.01035

Douglas, H. (2013). The Value of Cognitive Values. Philosophy of Science 80: 796-806.

Ellis, J., Gaillard, M.K. and Nanopoulos D.V., A phenomenological profile of the Higgs boson, Nucl. Phys. B 106 (1976) 292. (CERN preprint Nov. 1975)

Ellis, J., Gaillard, M.K., and Nanopoulos D.V. (2012). A Historical Profile of the Higgs Boson. arXiv:1201.6045v1 [hep-ph].

Ellis, J., Gaillard, M.K., and Nanopoulos D.V. (2015). An Updated Historical Profile of the Higgs Boson. http://arxiv.org/abs/1504.07217.

Franklin, A. (2013) Shifting Standards. Experiments in Particle Physics in the Twentieth Century. Pittsburgh: University of Pittsburgh Press.

Franklin, A. & Perovic, S. (2015). Experiment in Physics, *The Stanford Encyclopedia of Philosophy* (Fall 2013 Edition), Edward N. Zalta (ed.), URL = http://plato.stanford.edu/archives/sum2015/entries/physics-experiment/

Friederich, S., Harlander R., and Karaca, K. (2014). Philosophical perspectives on ad hoc-hypotheses and the Higgs mechanism. Synthese 191:3897–3917.

Giudice, G. (2008), Naturally Speaking: The Naturalness Criterion and Physics at the LHC, in G. Kane and A. Pierce (eds.), Perspectives on LHC Physics, Singapore: World Sciectific, 2008, pp. 155-178; also , arXiv:0801.2562.

Giudice, G. (2013), Naturalness after LHC8, PoS EPS-HEP2013 (2013) 163. 3, arXiv:1307.7879.

Grinbaum, A. (2012). Which fine-tuning arguments are fine? Foundations of Physics, 42(5), 615–631.

Gunion, J.F, Haber, H.E., Kane, G., Dawson, S. (1990). The Higgs Hunter's Guide. Redford City: Addison-Wesley.

INSPIRE https://inspirehep.net/info/general/project/index

Karaca, K (2013a). The strong and weak senses of theory-ladenness of experimentation: Theory-driven versus exploratory experiments in the history of high-energy elementary particle physics. Science in Context 26: 93-136.

Karaca, K. (2013b). The construction of the Higgs mechanism and the emergence of the electroweak theory. Studies in History and Philosophy of Modern Physics, 44 (1), 1-16.

Kuhn, T.S. (1977). Objectivity, Value Judgment, and Theory Choice. In: The Essential Tension. Chicago: Chicago University Press, pp. 320-329.

Lakatos, I. (1974) The Role of Crucial Experiments in Science. Studies in History and Philosophy of Science 4, 309-325.

Laudan, L. (1990). Demystifying Underdetermination. In C. Wade Savage, ed., *Scientific Theories* (267-297). Minneapolis: University of Minnesota Press.

Laudan, L., and J. Leplin. (1991). Empirical Evidence and Underdetermination. *Journal of Philosophy* 88: 449–472.

Mättig P (2019) Validation of Simulation in Particle Physics in: Beisbart, C. & Saam, N. J. (eds.), Computer Simulation Validation - Fundamental Concepts, Methodological Frameworks, and Philosophical Perspectives, Cham: Springer, to appear 2019

Norton, J. (2008). Must Evidence Underdetermine Theory? In M. Carrier, D. Howard, and J. Kourany, eds., *The Challenge of the Social and the Pressure of Practice: Science and Values Revisited* (pp. 17-44). Pittsburgh: University of Pittsburgh Press.

Peskin, M and Schroeder,S, (1995) An Introduction to Quantum Field Theory, Cambridge (Massachusetts): Perseus Books

Quigg, C. (2007), Spontaneous Symmetry Breaking as a Basis for Particle Masses, Rept.Prog.Phys. 70 (2007) 1019-1054, arXiv:0704.2232v2 [hep-ph]

Stanford, K. (2006) *Exceeding Our Grasp: Science, History, and the Problem of Unconceived Alternatives*, New York: Oxford University Press.

Stanford, Kyle, "Underdetermination of Scientific Theory", *The Stanford Encyclopedia of Philosophy* (Winter 2017 Edition), Edward N. Zalta (ed.), URL = <https://plato.stanford.edu/archives/win2017/entries/scientific-underdetermination/>.

Stöltzner, M. (2014) Higgs Models and Other Stories about Mass Generation. Journal for General Philosophy of Science 45, 369-384

Susskind, L. (1979). Dynamics of spontaneous symmetry breaking in the Weinberg–Salam theory. Physical ReviewD, 20(November), 2619–2625.

The Royal Swedish Academy of Science, The BEH-Mechanism, Interactions with Short Range Forces and Scalar Particles, 8.10.2013, http://www.nobelprize.org/nobel_prizes/physics/laureates/2013/advanced.html

't Hooft, Gerard (1979). Naturalness, chiral symmetry, and spontaneous chiral symmetry breaking. NATO Advanced Science Institutes Series B:Physics, 59 (PRINT-80-0083(UTRECHT)),135.

Weber, M. (2009). The Crux of Crucial Experiments: Duhem's Problems and Inference to the Best explanation. British Journal for the Philosophy of Science 60: 10-49.

Wells, J. (2012). Effective Field Theories and the Role of Consistency in Theory Choice, arXiv 1211.0634

Wells, J. (). Effective Theories in Physics. From Planetary Orbits to Elementary Particle Masses. Heidelberg: Springer 2012.

Wells, J. (2015). The utility of Naturalness, and how its application to Quantum Electrodynamics envisages the Standard Model and the Higgs boson. Studies in History and Philosophy of Modern Physics 49: 102-108.

Wells, J. (2017). Higgs naturalness and the scalar boson proliferation instability problem, Synthese 194, 477-490

Williams, P. (2015). Naturalness, the autonomy of scales, and the 125 GeV Higgs. Studies in History and Philosophy of Modern Physics 49: 102-108.

Wüthrich, A. (2016). The Higgs discovery as a diagnostic causal inference, Synthese 194, 461-476.